%% file: main.tex
\newif\ifdraft
\newif\ifwidemargins
\draftfalse          %
\ifdraft\widemarginstrue\fi
\PassOptionsToPackage{hyphens}{url}
\documentclass[camera,hyperref,letterpaper,nomarginnotes,nonarrowgutter]{jpaper}

\newcommand{\draftversion}{v3.0}
\providecommand{\githash}{unknown}
\providecommand{\buildhost}{unknown}

\usepackage{cite}

\usepackage{amsmath,amssymb,amsfonts}
\usepackage{algorithmic}
\usepackage{graphicx}
\usepackage{textcomp}
\usepackage{xcolor}

\usepackage{tikz}
\usepackage{subfig}
\usepackage{booktabs}
\usepackage{tabularx}
\usepackage{colortbl}
\usepackage{hhline}
\usepackage{multirow}
\usepackage{makecell}
\usepackage{balance}
\usepackage{mathtools}
\usepackage{array} %
\usepackage{siunitx} %
\usepackage{cancel}

\input{commands}

\input{ref_macros}
\input{sections/eval_params}

\selectfont

\usepackage{fancyhdr}
\ifdraft
\usepackage{datetime2}
\fancypagestyle{draftheader}{%
  \fancyhf{}%
  \fancyhead[C]{\textcolor{blue}{DRAFT \draftversion{} -- \today\ \DTMcurrenttime\ -- git commit: \githash\ -- \buildhost}}%
}
\fi

\def\BibTeX{{\rm B\kern-.05em{\sc i\kern-.025em b}\kern-.08em
    T\kern-.1667em\lower.7ex\hbox{E}\kern-.125emX}}

\begin{document}

\bstctlcite{IEEEexample:BSTcontrol}

\ifwidemargins
  \addtolength{\paperwidth}{4cm}
  \addtolength{\oddsidemargin}{2cm}
  \pdfpagewidth=\dimexpr 8.5in + 4cm\relax
  \makeatletter\let\marginpar\jp@marginpar\makeatother
  \setlength{\marginparwidth}{2cm}
  \setlength{\marginparsep}{2mm}
\else
  \pdfpagewidth=8.5in
\fi
\pdfpageheight=11in

\newcommand{\iscasubmissionnumber}{2768}

\pagenumbering{arabic}

\pretitle{\begin{center}\normalfont\fontsize{15}{18}\selectfont\bfseries}
\title{\papername: Efficient Solutions to the \CD{} Vulnerability\\ in DRAM-based Systems}
\author{
    Andreas Kosmas Kakolyris$^1$ \hspace{0.5em} F. Nisa Bostanc\i$^1$ \hspace{0.5em} Ataberk Olgun$^1$ \hspace{0.5em} \revi{\.{I}smail Emir Y{\"u}ksel}$^1$ \\
    Harsh Songara$^1$ \hspace{0.5em} Konstantinos Marios Sgouras$^1$ \hspace{0.5em}  Umut Ba\c{s}er$^{1,2}$ \\
    Konstantinos Kanellopoulos$^1$ \hspace{0.5em} A. Giray Ya\u{g}l\i k\c{c}\i$^3$ \hspace{0.5em} Onur Mutlu$^1$ \vspace{0.25em} \\
    \normalsize{
        $^1$ETH Z{\"u}rich \hspace{0.5em} $^2$\revi{TOBB ET{\"U}} \hspace{0.5em} $^3$CISPA
    }
}

\maketitle
\ifdraft
\thispagestyle{draftheader}
\pagestyle{draftheader}
\else
\pdfpagewidth=8.5in
\pdfpageheight=11in
\thispagestyle{plain}
\pagestyle{plain}
\fi

\input{sections/00_abstract}
\input{sections/01_intro}

\input{sections/02_background}

\input{sections/03_mechanism}
\input{sections/04_security}
\input{sections/05_methodology}
\input{sections/06_evaluation}
\input{sections/07_implementation}
\input{sections/08_indram}
\input{sections/09_discussion}
\input{sections/10_related}
\input{sections/11_conclusion}
\input{sections/12_acknowledgements}

\balance
\bibliographystyle{IEEEtran}
\bibliography{refs}

\clearpage
\input{sections/13_artifact_appendix}

\end{document}

%% file: commands.tex
\newcommand{\papername}{ColumnKeeper}
\newcommand{\CD}{ColumnDisturb}
\newcommand{\RH}{RowHammer}
\newcommand{\CGUARDD}{\papername-D}
\newcommand{\CGUARDP}{\papername-P}

\newcommand{\CKD}{CK-D}
\newcommand{\CKP}[1][]{CK-P\if\relax\detokenize{#1}\relax\else#1\fi}
\newcommand{\CKS}{CK-S}
\newcommand{\CKDSALP}{SALP-CK-D}
\newcommand{\CKPSALP}{SALP-CK-P}

\newcommand{\secref}[1]{§\ref{#1}}

\newcommand{\cdcounter}[2]{$c^{#2}_{#1}$}
\newcommand{\nth}{$N_{CD}$}

\newcommand{\CTO}{CT-O}
\newcommand{\CTE}{CT-E}
\newcommand{\RPT}{RPT}

\definecolor{nbs}{rgb}{0.88, 0.07, 0.37}
\definecolor{ao}{rgb}{0.67, 0.31, 0.69}
\definecolor{iy}{rgb}{0.0, 0.36, 0.05}
\definecolor{gfored}{rgb}{0.580, 0.050, 0.211}
\definecolor{omcolor}{rgb}{1.0, 0.0, 0.0}

\definecolor{revbluegreen}{HTML}{0E8C7B}
\definecolor{watermelon}{HTML}{FC6C85}
\definecolor{akklightblue}{HTML}{E6F0FF}
\newcommand{\revi}[1]{#1}
\newcommand{\revii}[1]{#1}
\newcommand{\reviii}[1]{#1}
\newcommand{\reviv}[1]{#1}

\usepackage[colorinlistoftodos,prependcaption,textsize=tiny]{todonotes}
\usepackage{color,soul}

\ifdraft
\newcommand{\param}[1]{\textcolor{red}{#1}}

\newcommand{\agycomment}[1]{\todo[size=\scriptsize, linecolor=orange, bordercolor=orange, backgroundcolor=white]{\textcolor{gfored}{\textbf{@gy:} #1}}}

\newcommand{\nbcomment}[1]{\todo[size=\scriptsize, linecolor=orange, bordercolor=orange, backgroundcolor=white]{\textcolor{nbs}{\textbf{@nb:} #1}}}

\newcommand{\ieycomment}[1]{\todo[size=\scriptsize, linecolor=orange, bordercolor=orange, backgroundcolor=white]{\textcolor{iy}{\textbf{@iey:} #1}}}

\newcommand{\atbcomment}[1]{\todo[size=\scriptsize, linecolor=orange, bordercolor=orange, backgroundcolor=white]{\textcolor{ao}{\textbf{@atb:} #1}}}

\newcommand{\akkcomment}[1]{}

\newcommand{\om}[1]{\todo[size=\normalsize, linecolor=red, bordercolor=red, backgroundcolor=yellow]{\textcolor{omcolor}{\textbf{\Large @OM:\ #1}}}}

\newcommand{\revv}[1]{\textcolor{watermelon}{#1}}

\else
    
    \newcommand{\agycomment}[1]{}
    
    \newcommand{\atbcomment}[1]{}
    
    \newcommand{\akkcomment}[1]{}
    \newcommand{\om}[1]{}
    
    \newcommand{\nbcomment}[1]{}
    
    \newcommand{\ieycomment}[1]{}
    \newcommand{\param}[1]{{#1}}
    
    \newcommand{\revv}[1]{#1}
\fi

\newcommand*\circled[1]{(\tikz[baseline=(char.base)]{\node[shape=circle,fill,inner sep=0.5pt] (char) {\footnotesize\textcolor{white}{#1}};})}

%% file: ref_macros.tex
\newcommand{\exploitingRowHammerAllCitations}[0]{\cite{fournaris2017exploiting, poddebniak2018attacking, tatar2018throwhammer, carre2018openssl, barenghi2018software, zhang2018triggering, bhattacharya2018advanced, kim2014flipping, rowhammergithub, seaborn2015exploiting, van2016drammer, gruss2016rowhammer, razavi2016flip, pessl2016drama, xiao2016one, bosman2016dedup, bhattacharya2016curious, burleson2016invited, qiao2016new, brasser2017can, jang2017sgx, aga2017good, mutlu2017rowhammer, tatar2018defeating, gruss2018another, lipp2018nethammer, van2018guardion, frigo2018grand, cojocar2019eccploit,  ji2019pinpoint, mutlu2019rowhammer, hong2019terminal, kwong2020rambleed, frigo2020trrespass, cojocar2020rowhammer, weissman2020jackhammer, zhang2020pthammer, yao2020deephammer, deridder2021smash, hassan2021utrr, jattke2022blacksmith, tol2022toward, kogler2022half, orosa2022spyhammer, zhang2022implicit, liu2022generating, cohen2022hammerscope, zheng2022trojvit, fahr2022frodo, tobah2022spechammer, rakin2022deepsteal, kang2024sledgehammer, lin2025gpuhammer, nazaraliyev2025not, meyer2026phoenix,kamadan2025ecc}}

\newcommand{\mitigatingRowHammerAllCitations}[0]{\cite{%
lee2019twice, seyedzadeh2018cbt, kang2020cattwo, park2020graphene, kim2022mithril, kim2014architectural, bains2015row, bains2016distributed, bains2016row, bostanci2024comet, canpolat2024understanding, canpolat2025chronus, jedecddr5c, bennett2021panopticon,olgun2024abacus, yaglikci2024svard,qureshi2026salt,%
yaglikci2021blockhammer, canpolat2024breakhammer, greenfield2012throttling,%
saxena2022aqua,%
saileshwar2022randomized, kim2014flipping, yaglikci2022hira, you2019mrloc, son2017making, vittal2025mopac, qureshi2024mint,taneja2026mirza,wu2026apt,%
AppleRefInc, rh-hp, rh-lenovo, bains14d, bains14c, bains14a, bains14b, aweke2016anvil, irazoqui2016mascat, yaglikci2021security, frigo2020trrespass, hassan2021utrr, qureshi2022hydra, brasser2017can, konoth2018zebram, van2018guardion, vig2018rapid, lee2021cryoguard, marazzi2022protrr, zhang2022softtrr, joardar2022learning, juffinger2023csi, enomoto2022efficient, manzhosov2022revisiting, ajorpaz2022evax, naseredini2022alarm, joardar2022machine, hassan2024self, zhang2020leveraging, loughlin2021stop, devaux2021method, fakhrzadehgan2022safeguard, saroiu2022price, loughlin2022moesiprime, han2021surround, mutlu2023fundamentally, woo2023scalable, bock2019riprh, wang2021discreet, taneja2025dream, lin2025cnc, qazi2025drfm, lu2025counterpoint, qureshi2025autorfm, woo2025dapper, woo2025qprac,wi2026rowarmor%
}}

\newcommand{\deterministicRowHammerDefenseCitations}[0]{\cite{lee2019twice, yaglikci2021blockhammer, seyedzadeh2017cbt, seyedzadeh2018cbt, kang2020cattwo, park2020graphene, kim2022mithril, saileshwar2022randomized, saxena2022aqua, kim2014architectural, bains2015row, bains2016distributed, bains2016row,bostanci2024comet,canpolat2025chronus,olgun2024abacus,qureshi2026salt}}

\newcommand{\probabilisticRowHammerDefenseCitations}[0]{\cite{kim2014flipping, yaglikci2022hira, you2019mrloc, son2017making,vittal2025mopac,qureshi2024mint,taneja2026mirza,wu2026apt}}

\newcommand{\refreshBasedRowHammerDefenseCitations}[0]{\cite{lee2019twice, seyedzadeh2017cbt, seyedzadeh2018cbt, kang2020cattwo, park2020graphene, kim2022mithril, kim2014architectural, bains2015row, bains2016distributed, bains2016row, aweke2016anvil,olgun2024abacus,bostanci2024comet,canpolat2024understanding,canpolat2025chronus,qureshi2026salt,taneja2026mirza,wu2026apt}}

\newcommand{\rowHammerGetsWorseCitations}[0]{\cite{kim2014flipping, kim2020revisiting, frigo2020trrespass, yaglikci2022understanding, orosa2021deeper, mutlu2017rowhammer, mutlu2018rowhammer, mutlu2019rowhammer, mutlu2023fundamentally}}

\newcommand{\ramulatorCitations}[0]{\cite{kim2016ramulator, luo2023ramulator2, ramulator2github,ramulatorgithub, bostanci2026cleaning}}

\newcommand{\pracCitations}[0]{\cite{jedecddr5c,canpolat2024understanding,canpolat2025chronus,woo2025qprac,bennett2021panopticon,kim2023ddr5}}

\newcommand{\RHFundamentalCitations}[0]{\cite{kim2014flipping,mutlu2023retrospective,mutlu2017rowhammer,mutlu2019rowhammer,cojocar2020rowhammer,mutlu2023fundamentally}}

\newcommand{\rowPressCitations}[0]{\cite{luo2023rowpress,luo2024rowpress}}

\newcommand{\RHandRP}[0]{\RH{}~\RHFundamentalCitations{} and RowPress~\rowPressCitations{}}

\newcommand{\salpCitations}[0]{\cite{kim2012case,chang2014improving,zhang2014cream,yaglikci2022hira}}

\newcommand{\openBitlineCitations}[0]{\cite{chang2016lisa,itoh77,jacob_book_2008,luo2020clrdram,sekiguchi2002low,columndisturb,yuksel2024functionally}}

\newcommand{\readDisturbanceCitations}[0]{\cite{kim2014flipping,mutlu2023retrospective,mutlu2017rowhammer,mutlu2019rowhammer,cojocar2020rowhammer,mutlu2023fundamentally,luo2023rowpress,luo2024rowpress,columndisturb,cai-dsn15}}

%% file: sections/eval_params.tex
\newcommand{\IpcSCMinDHi}{0.98}
\newcommand{\IpcSCMinPaHi}{0.97}
\newcommand{\IpcSCMinPbHi}{0.96}
\newcommand{\SlSCAvgDHi}{0.15}
\newcommand{\SlSCAvgPbHi}{0.36}
\newcommand{\IpcSCGmDMi}{0.98}
\newcommand{\IpcSCGmPaMi}{0.97}
\newcommand{\IpcSCGmPbMi}{0.97}
\newcommand{\SlSCAvgDMi}{1.70}
\newcommand{\SlSCAvgPbMi}{2.73}
\newcommand{\IpcSCGupsPaMi}{0.78}
\newcommand{\IpcSCGupsPbMi}{0.76}
\newcommand{\IpcSCGmDLo}{0.84}
\newcommand{\IpcSCGmPaLo}{0.79}
\newcommand{\IpcSCGmPbLo}{0.78}
\newcommand{\MrSCRatioPaOverDLo}{1.70}
\newcommand{\MrSCRatioPbOverDLo}{1.83}

\newcommand{\EnSCAvgDMi}{1.45}
\newcommand{\EnSCMaxDMi}{9.04}
\newcommand{\EnSCAvgPaMi}{2.19}
\newcommand{\EnSCMaxPaMi}{17.82}
\newcommand{\EnSCAvgPbMi}{2.36}
\newcommand{\EnSCMaxPbMi}{19.08}
\newcommand{\EnSCAvgDLo}{16.26}
\newcommand{\EnSCAvgPaLo}{22.82}
\newcommand{\EnSCAvgPbLo}{24.69}
\newcommand{\EnSCHighMaxDLo}{2.00}
\newcommand{\EnSCHighMaxPaLo}{3.14}
\newcommand{\EnSCHighMaxPbLo}{3.33}

\newcommand{\SlMCAvgDMi}{6.56}
\newcommand{\SlMCAvgPaMi}{9.52}
\newcommand{\SlMCAvgPbMi}{10.17}
\newcommand{\SlMCDvsPbMi}{3.61}
\newcommand{\SlMCDvsPbLo}{11.13}
\newcommand{\SlSCDvsPbMi}{1.04}
\newcommand{\SlSCDvsPbLo}{6.19}

\newcommand{\EnMCAvgDMi}{5.95}
\newcommand{\EnMCMaxDMi}{11.27}
\newcommand{\EnMCAvgPaMi}{8.83}
\newcommand{\EnMCAvgPbMi}{9.54}
\newcommand{\EnMCAvgDLo}{1.72}
\newcommand{\EnMCAvgPaLo}{2.05}
\newcommand{\EnMCAvgPbLo}{2.15}

\newcommand{\IpcAdvAllHiMaxOh}{0.47}
\newcommand{\IpcAdvAllMiMaxOh}{3.68}
\newcommand{\IpcAdvAllLoMinOh}{20.97}
\newcommand{\MrAdvRatioDLoOverMi}{10.47}
\newcommand{\MrAdvRatioSLoOverMi}{10.92}
\newcommand{\MrAdvRatioPLoOverMi}{11.08}
\newcommand{\MrAdvSvsDLo}{5.06}
\newcommand{\MrAdvSvsPbLo}{27.50}
\newcommand{\IpcAdvAvgDLo}{20.97}
\newcommand{\IpcAdvAvgSLo}{21.52}
\newcommand{\IpcAdvAvgPbLo}{29.07}

\newcommand{\IpcRhGrapheneAvgD}{0.97}
\newcommand{\IpcRhGrapheneMinD}{0.78}
\newcommand{\IpcRhCdOnlyAvgD}{0.98}
\newcommand{\IpcRhCdOnlyMinD}{0.88}
\newcommand{\SlRhPracAvgD}{1.70}
\newcommand{\SlRhPracAvgPb}{2.61}
\newcommand{\SlRhHydraAvgD}{2.15}
\newcommand{\SlRhHydraAvgPb}{3.08}

\newcommand{\IpcFragGmDTwenty}{1.009}
\newcommand{\IpcFragGmDFifty}{1.008}
\newcommand{\IpcFragGmDEighty}{1.006}
\newcommand{\IpcFragGmPbTwenty}{1.014}
\newcommand{\IpcFragGmPbFifty}{1.012}
\newcommand{\IpcFragGmPbEighty}{1.009}

\newcommand{\SlSubAvgDOneKMi}{1.70}
\newcommand{\SlSubAvgDTwoFiveSixMi}{0.31}
\newcommand{\SlSubAvgPbOneKMi}{2.71}
\newcommand{\SlSubAvgPbTwoFiveSixMi}{0.92}
\newcommand{\SlSubAvgDSmallLo}{1.64}
\newcommand{\SlSubMaxDSmallLo}{12.57}
\newcommand{\SlSubAvgPbSmallLo}{4.34}
\newcommand{\SlSubMaxPbSmallLo}{34.18}

\newcommand{\SlSalpAvgDHi}{0.11}
\newcommand{\SlSalpMaxDHi}{1.43}
\newcommand{\SlSalpAvgPbHi}{0.16}
\newcommand{\SlSalpMaxPbHi}{1.92}
\newcommand{\SlSalpAvgDMi}{0.87}
\newcommand{\SlSalpMaxDMi}{10.71}
\newcommand{\SlSalpAvgPbMi}{1.20}
\newcommand{\SlSalpMaxPbMi}{14.00}
\newcommand{\SlSalpAvgDLo}{7.87}
\newcommand{\SlSalpMaxDLo}{56.42}
\newcommand{\SlSalpAvgPbLo}{10.62}
\newcommand{\SlSalpMaxPbLo}{65.34}
\newcommand{\SlNoSalpMaxDLo}{61.79}
\newcommand{\SlNoSalpMaxPbLo}{79.16}

\newcommand{\DiscIpcInDramAvgLo}{0.97}
\newcommand{\DiscIpcInDramMinLo}{0.84}
\newcommand{\DiscIpcInDramMinMi}{0.99}
\newcommand{\DiscIpcMemCtrAvgLo}{0.84}
\newcommand{\DiscIpcMemCtrMinLo}{0.41}

%% file: sections/00_abstract.tex
\begin{abstract}
Modern DRAM chips are vulnerable to read disturbance phenomena such as RowHammer and RowPress, which induce bitflips in DRAM rows after accessing nearby rows a certain number of times (i.e., the read disturbance threshold).
\CD{} \revi{is} a new and fundamentally different DRAM read disturbance phenomenon.
Specifically, \CD{} (i)~disturbs DRAM columns instead of rows, and (ii)~increases the number of affected DRAM cells from those that reside in \revii{only} a few neighboring rows to \emph{all} cells across three consecutive DRAM subarrays.%

We propose \papername, the first \revii{set of \CD{} mitigations} that \revii{has} two variants: \CGUARDD{} (\CKD), a deterministic mitigation mechanism, and \CGUARDP{} (\CKP), a probabilistic one.
The key idea of \CKD{} is to take advantage of DRAM's \textit{open-bitline} architecture to provide deterministic security guarantees against \CD{} with low performance and energy overheads. 
To achieve this, \CKD{} employs two counters per subarray to track the number of activations affecting the odd and even columns of the subarray, respectively. 
When either counter reaches a predetermined threshold, \CKD{} refreshes \revi{one} row in the \revii{corresponding} subarray.
The key idea of \CKP{} is to refresh \revi{one row in three consecutive subarrays upon a row activation in the \emph{middle} subarray.}
This occurs with a predetermined probability and provides configurable security guarantees at low area overhead. 
To identify which row to refresh in each subarray, both mechanisms maintain a table with one entry per subarray that always points to the next row to refresh within the subarray.
Each time either mechanism refreshes a row, it increments the corresponding entry, \revi{thereby} ensuring that all rows in the subarray are refreshed in a round-robin manner.

Our comprehensive evaluation shows that both mechanisms prevent \CD{} bitflips at low performance, energy, and area overheads. 
At the current \revi{experimentally-demonstrated} \CD{} threshold (1M), \CKD{} and \CKP{} incur very low average single-core performance overheads of \param{\SlSCAvgDHi\%} and \param{\SlSCAvgPbHi\%}, respectively, compared to a system with no \CD{} \revi{mitigation}.
For near-future thresholds (128K), these overheads rise to a still low average of \param{\SlSCAvgDMi\%} and \param{\SlSCAvgPbMi\%}, respectively.
Our experimental analysis shows that mitigating \CD{} with low performance overheads at low thresholds (e.g., 16K) is still possible by either adopting smaller subarray sizes or enabling subarray-level parallelism. 
\CKD{} and \CKP{} can be implemented with low area overheads of 0.1$\text{mm}^2$ and 0.03$\text{mm}^2$, respectively.
\revii{\papername{} is freely available at \revi{\href{https://github.com/CMU-SAFARI/ColumnKeeper}{github.com/CMU-SAFARI/ColumnKeeper}}.}

\end{abstract}

%% file: sections/01_intro.tex
\section{Introduction}

Modern DRAM chips are susceptible to read disturbance phenomena such as \RHandRP{}.
These phenomena can induce bitflips in DRAM cells located in a ``victim'' row by \reviii{repeatedly} (i) activating and precharging (i.e., ``hammering'') neighboring ``aggressor'' rows, or (ii) keeping an aggressor row open for long periods of time (i.e., ``pressing'').
By doing so, such phenomena break memory isolation, i.e., the assumption that accessing a specific memory address does not affect data stored in other memory addresses. 
As DRAM technology node scaling progresses, the severity of \RH{} and RowPress increases~\rowHammerGetsWorseCitations{}. 
Specifically: (i) the number of \revii{row activations} required to induce bitflips (\reviii{i.e.,} the read disturbance threshold) decreases~\cite{mutlu2023fundamentally,kim2014flipping,kim2020revisiting,cojocar2020rowhammer,orosa2021deeper}, and (ii) the number of experienced bitflips increases~\cite{loughlin2022moesiprime,kim2020revisiting}.
Prior works~\cite{frigo2020trrespass,tatar2018throwhammer,gruss2016rowhammer,seaborn2015exploiting,poddebniak2018attacking,fournaris2017exploiting} show that attackers can leverage read disturbance bitflips in real systems~\exploitingRowHammerAllCitations{} for a wide variety of attacks, such as (i) privilege escalation, (ii) \reviii{leaking sensitive information}, (iii) \reviii{virtual machine escapes}, (iv) \reviii{data corruption}, and (v) denial-of-service.
To ensure system robustness (i.e., safety, security, reliability, and availability), prior works~\mitigatingRowHammerAllCitations{} propose various mitigation mechanisms that often work by probabilistically or deterministically tracking aggressor row activations and refreshing potential victim rows before they experience bitflips.

\reviii{\emph{In contrast}} to row-based read disturbance phenomena~\cite{kim2014flipping,luo2023rowpress}, the newly discovered \reviii{\emph{column-based}} read disturbance phenomenon~\cite{columndisturb}, \reviii{\emph{\CD{}}}, occurs by repeatedly ``hammering'' the same \emph{column} of DRAM cells.
As a result, \CD{} affects all DRAM cells that share a bitline (i.e., cells in the same physical DRAM column).
Due to the open-bitline architecture of modern DRAM~\openBitlineCitations{} in which bitlines are shared between neighboring subarrays (\secref{sub:dram_org}), \CD~\cite{columndisturb} induces bitflips in \emph{three} \revii{consecutive} subarrays.
As a result, each hammer affects thousands of rows and millions of cells at once~\cite{columndisturb}.
\reviii{As such,} existing \RH{} mitigation mechanisms~\mitigatingRowHammerAllCitations{} are incapable of defending against \CD{} \revii{because they} (i) \revii{can prevent} bitflips \reviii{\emph{only in a few}} rows adjacent to the aggressor row, and (ii) track activations at row granularity~\cite{qureshi2025moat,vittal2025mopac,canpolat2025chronus}.

\noindent\textbf{Our goal} in this work is to design the first \reviii{practical} mechanisms that mitigate \CD{} bitflips at current and future \CD{} thresholds.
To this end, we introduce \papername{},\footnote{Similarly to how a goalkeeper~\cite{wiki:goalkeeper} prevents an opposing team from scoring in the game of football, \papername{} prevents \CD{} bitflips.} a set of mitigation mechanisms that prevent \CD{} bitflips at low performance, energy, and area overheads.
\papername{} \revii{has} two variants: (i) \CGUARDD{}, a deterministic mitigation mechanism, and (ii) \CGUARDP{}, a probabilistic one.

\revii{\noindent\textbf{Key Ideas.}
The overarching idea behind both \papername{} variants is to track activations at the \reviii{\emph{subarray granularity}} and iteratively refresh \emph{all} rows in affected subarrays before \CD{} bitflips can occur.
The two variants differ in \reviii{\emph{how}} they track these activations.}

\revii{The key idea \reviii{of} \CGUARDD{} (\CKD) is to account for DRAM's open-bitline architecture to accurately count the number of activations in each subarray.
Specifically, we observe that each activation hammers \emph{both} the odd and the even columns in the activated subarray, but either the odd \emph{or} the even columns in \reviii{the two adjacent} neighbor \reviii{subarrays, respectively}.
\CKD{} leverages this observation and separately counts activations for the odd and even columns of subarrays to avoid overestimating activations and issuing unnecessary refreshes.}

\revii{The key idea \reviii{of} \CGUARDP{} (\CKP) is to probabilistically issue preventive refreshes following each row activation.
This eliminates the need for activation counting and provides configurable security guarantees at lower hardware overheads.}

\revii{\noindent\textbf{Key Mechanisms.}
To count activations, \CKD{} maintains two counter tables: \emph{Counter Table-Even} (\CTE) and \emph{Counter Table-Odd} (\CTO), each containing one entry per subarray.
Upon a row activation, \CKD{} increments both the \CTE{} and \CTO{} entries for the activated subarray, but \emph{only} one of them for \reviii{each of the} neighboring subarrays.
Each time the \CTO{} or \CTE{} entry of a subarray reaches a predetermined threshold, \CKD{} issues a preventive refresh to the corresponding subarray.}

\revii{\CKP{} \reviii{avoids} activation counting altogether. Instead, upon a row activation in a subarray, it issues preventive refreshes to the corresponding subarray and its neighboring subarrays with a predetermined probability.}

\revii{To identify \emph{which} rows to refresh within a subarray, both \CKD{} and \CKP{} employ a \emph{Row Pointer Table} (\RPT) that stores one entry per subarray, which points to the next row to refresh in each subarray.
On each preventive refresh, both variants first probe and then increment the relevant \RPT{} entry, so that all rows in affected subarrays are iteratively refreshed in a round-robin manner before \CD{} bitflips can occur.}

\noindent\textbf{Key Results.}
We evaluate the impact of \papername's two \revii{variants} on system performance and energy efficiency using Ramulator 2.0~\ramulatorCitations{} and DRAMPower~\cite{drampower,steiner2025drampower} across 62 single- and 60 multi-core workloads (see \secref{sec:methodology}). 
We observe that at what we consider a current \CD{} threshold (\nth) of $1M$, \CKD{} and \reviii{\CKP{}} incur very low average performance overheads of \param{\SlSCAvgDHi\%} and \param{\SlSCAvgPbHi\%}, respectively, across single-core workloads, compared to a baseline with no \CD{} mitigation.
At a ``\textit{near-future}'' threshold of $128K$, the performance overheads increase to a still low average of \param{\SlSCAvgDMi\%} and \param{\SlSCAvgPbMi\%}, respectively, \reviii{across single-core workloads}.
We show that mitigating \CD{} for a very low threshold of $16K$ is still possible at low overheads, by (i) reducing subarray sizes, (ii) enabling subarray-level parallelism (SALP)~\salpCitations{}, or (iii) with a potential in-DRAM implementation of \CKD{} (\secref{disc:indram}).

Our contributions in this paper are as follows:
\begin{itemize}
    \item We present \papername, \revii{the first} set of mechanisms that protect against \CD{} bitflips at low performance, energy, and hardware overheads.
    
    \item We propose \CGUARDD{} (\CKD), a deterministic mitigation mechanism whose key idea is to separately count activations for the odd and even columns of subarrays.
    \revii{\reviii{Doing so} accounts for modern DRAM's open-bitline architecture and avoids overestimating activations, \reviii{thereby} preventing unnecessary refreshes and incurring low performance and energy overheads.} 
    \CKD{} issues preventive refreshes each time the activation count of either the even or the odd columns \reviii{of a subarray} reaches a predetermined threshold.
    
    \item We propose \CGUARDP{} (\CKP), a probabilistic mechanism with lower hardware overheads.
    The key idea \reviii{of} \CKP{} is to avoid counting activations \revii{and instead issue preventive refreshes to one row in three consecutive subarrays} with a predetermined probability. \reviii{\CKP{}} provides configurable security guarantees at lower hardware overheads. 

    \item \revii{We analyze the security of both mechanisms, showing that: (i) \CKD{} deterministically prevents \CD{} bitflips \reviii{against worst-case access patterns} (\secref{security:d}), and (ii) \CKP{} guarantees a configurable, arbitrarily low probability of experiencing \CD{} bitflips within a given period of time (e.g., a year; see \secref{security:p} and \secref{sec:montecarlo}).}

    \item We comprehensively evaluate the security, performance, energy, and area overheads of both mitigation mechanisms. 
    We show that for current thresholds ($1M$), both mechanisms incur ${<}1\%$ performance overheads, on average, while for ``\textit{near-future}'' thresholds, \revii{overheads} rise to ${<}3\%$. 
    We show that with smaller subarray sizes, subarray-level parallelism, or a potential in-DRAM implementation, \papername{} can mitigate \CD{} with low overheads even \reviii{at a very low} threshold of $16K$.

\end{itemize}

%% file: sections/02_background.tex
\section{Background \& Motivation}
\label{sec:background}

\subsection{DRAM Organization and Operation}
\label{sub:dram_org}

\subsubsection{DRAM Organization}

\begin{figure}[ht]
\centering
        \includegraphics[width=\columnwidth]{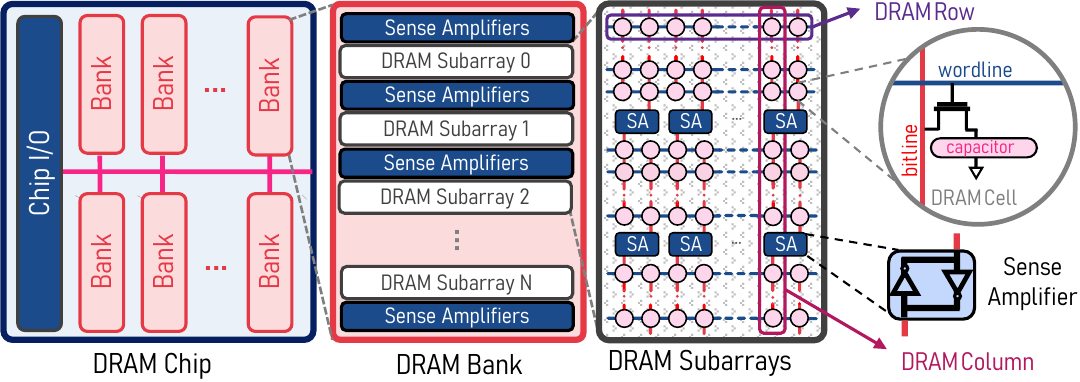}
        \caption{DRAM chip, bank, \& subarray organization.}
        \label{fig:dram_organization}
\end{figure}

Modern computing systems use DRAM as main memory, organizing it into a hierarchy with the following levels: channels, modules, ranks, banks, and subarrays.
Figure~\ref{fig:dram_organization} shows the lower levels of this hierarchy.
In CPUs, one or more memory controllers each interface with a DRAM channel to serve memory requests.
The DRAM channel contains one or more DRAM modules, which comprise multiple DRAM ranks, that is, a number of DRAM chips operating in lockstep.
Internally, DRAM chips contain multiple banks.
A bank contains multiple subarrays~\cite{seshadri2013rowclone,kim2012case,yuksel2024functionally,chang2016lisa} that each form a 2D array of DRAM cells organized into columns and rows.
A DRAM cell stores a single bit of information in the form of charge stored in a capacitor.
A voltage \revv{level} of $V_{DD}$ ($GND$) represents a logical 1 (0).
To access the DRAM cell, an access transistor connects the cell to a bitline, which is shared among all cells in the same DRAM column.
The access transistor is controlled by a \textit{wordline} that is shared between cells located in the same row.
When the wordline is asserted, an access transistor connects each cell of the row to its corresponding bitline, whose voltage is perturbed.
A structure called the \textit{row buffer}, which consists of an array of sense amplifiers (SAs), senses and amplifies this perturbation.
Due to a large size discrepancy between cells and SAs, modern DRAM adopts the \textit{open-bitline architecture}~\openBitlineCitations{}, which shares SAs between neighboring subarrays by connecting their bitlines to a shared SA in an interleaved manner.
Each bitline in a subarray connects to an SA that is also connected to a bitline in the subarray above or below.

\subsubsection{DRAM Operation}
The memory controller issues DRAM commands that: (i)~activate rows (\texttt{ACT}), (ii)~precharge bitlines (\texttt{PRE}) to $V_{DD}{/}2$, (iii)~read/write data (\texttt{RD/WR}), and (iv)~refresh DRAM rows (\texttt{REF}).
To perform DRAM reads or writes, the memory controller first \textit{opens} a DRAM row by issuing \texttt{ACT}, which latches \revv{the row's} data \revv{into} the row buffer.
To read (write) data, the memory controller then sends \texttt{RD} (\texttt{WR}) commands.
The scheduling of DRAM commands is governed by DRAM timing constraints~\cite{kim2012case,lee2013tiered,lee2015adaptive} such as $t_{RAS}$ (i.e., the minimum time between opening and closing a row) and $t_{RC}$ (i.e., the minimum time between issuing \texttt{ACT} to two different rows in the same bank).
Accesses to already open rows are served faster by omitting the \texttt{ACT} command, while accesses to different rows are slower since the bitlines must first be precharged via a \texttt{PRE} command (also referred to as \textit{closing} a row), before issuing \texttt{ACT}.

\subsubsection{Periodic DRAM Refresh} Over time, charge leaks from a DRAM cell's capacitor~\cite{liu2012raidr,mutlu2023retrospective-raidr,qureshi2015avatar,liu2013experimental}.
If left unchecked, this eventually results in \textit{retention failure}, i.e., bitflips due to the voltage reaching levels that cannot be reliably sensed by the SAs.
To ensure robust operation, the memory controller periodically issues \texttt{REF} commands that restore the charge of all cells in a row by opening and then closing it.
The time between two subsequent refreshes of the same row, called the \textit{refresh window} ($t_{REFW}$), is usually $32\,\text{ms}$ and $64\,\text{ms}$ for DDR5~\cite{jedecddr5c} and DDR4~\cite{jedec2017ddr4}, respectively.
In order to refresh \emph{all} rows within the \textit{refresh window}, the memory controller issues \texttt{REF} commands at every \textit{refresh interval} ($t_{REFI}$), which is typically $3.9\,\mu\text{s}$ for DDR5 and $7.8\,\mu\text{s}$ for DDR4.

\subsection{DRAM Read Disturbance}
\label{sub:dram_read_disturbance}
Apart from \textit{retention failure}, DRAM is also vulnerable to \textit{read disturbance}, i.e., the phenomenon where reading data (from DRAM or storage) causes data corruption due to the physical properties of the storage medium~\readDisturbanceCitations{}.
DRAM is known to be vulnerable to three types of such phenomena: \RH{}~\RHFundamentalCitations{}, RowPress~\rowPressCitations{}, and \CD~\cite{columndisturb}.

\noindent\textbf{\RH{} and RowPress} are row-based read disturbance phenomena that occur when rows are repeatedly accessed (hammered) or kept open (pressed) for long periods of time.
These phenomena cause bitflips in \emph{physically nearby} rows, and their effects manifest \emph{up to a few rows away} (typically 2-4 rows~\cite{lang2023blaster,kogler2022half,kim2020revisiting}), which is referred to as the \textit{blast radius}.
Inducing \RH{} or RowPress \revv{bitflips} requires hammering or pressing rows a certain number of times (i.e., the read disturbance threshold, \revv{$N_{RH}$} for \RH{} and $N_{RP}$ for RowPress).
Aggressive technology node scaling reduces this threshold, with $N_{RH}$ dropping from 139K hammers in~\cite{kim2014flipping} to only 4.8K hammers in 2020~\cite{kim2020revisiting}.
\revv{Similarly, in newer DRAM chips, \emph{all} DRAM rows are RowPress-vulnerable (compared to $\approx$60\% in older chips from the same manufacturer)~\cite{luo2023rowpress,luo2024rowpress}.}

\noindent\textbf{\CD{}} is a new, fundamentally different, column-based read disturbance phenomenon where hammering or pressing rows induces bitflips in DRAM cells sharing the same \emph{column} (i.e., the same physical DRAM bitline).
Due to the open-bitline architecture~\openBitlineCitations{} of modern DRAM (see \secref{sub:dram_org}), a subarray $k$ \revv{shares half of its bitlines with the subarray above it (i.e., $k{-}1$) and half with the subarray below it (i.e., $k{+}1$) in an interleaved fashion}.
\revv{As a result, each row activation in subarray $k$ affects (i) \emph{all} cells in subarray $k$, (ii) \emph{all} cells residing in \emph{even} columns of subarray $k{-}1$, and (iii) \emph{all} cells residing in \emph{odd} columns of subarray $k{+}1$ (or vice versa)}.
\revv{For a subarray size of 1024 rows as reported in~\cite{columndisturb}, the blast radius of \CD{} spans up to 3072 \revv{rows} in total, with each hammer affecting millions of cells.}
\CD{} has been shown to induce bitflips in $63.6\,\text{ms}$~\cite{columndisturb}, which falls within the DDR4 refresh window ($t_{REFW}$) of $64\,\text{ms}$~\cite{jedec2017ddr4}.
\revv{By dividing the minimum time to induce \CD{} bitflips ($63.6\,\text{ms}$) by $t_{RC}{\approx}50\,\text{ns}$ (i.e., the minimum time between two successive row activations in the same bank, see \secref{sub:dram_org}),
we calculate that the \emph{maximum} number of row activations required to induce \CD{} bitflips (i.e., the \CD{} threshold, $N_{CD}$) is $\approx1.2M$ activations.}
\revv{We conservatively round this number down to the nearest power of two (i.e., $1M$) to define the current $N_{CD}$.}
As the severity of read disturbance increases with aggressive technology node scaling~\rowHammerGetsWorseCitations{}, \CD{} has significant implications for the \revv{robustness} and performance of future DRAM-based computing systems.

\subsubsection{Read Disturbance Mitigation Techniques}
\label{sub:read_disturbance_mitigation}

To ensure robust operation, prior works propose a wide range of techniques that mitigate the effects of read disturbance~\mitigatingRowHammerAllCitations{}, \revv{particularly \RH{}~\RHFundamentalCitations{} and more recently RowPress~\cite{luo2023rowpress,luo2024rowpress}.}
These mitigation mechanisms combine architectural insights and algorithmic techniques to track DRAM row activations and deploy appropriate countermeasures, i.e., mechanisms that prevent \RH{} bitflips.
Tracking mechanisms exist on a spectrum spanning purely probabilistic (e.g.,~\probabilisticRowHammerDefenseCitations), deterministic (e.g.,~\deterministicRowHammerDefenseCitations), and hybrid (e.g.,~\cite{olgun2024abacus,bostanci2024comet,park2020graphene}) techniques.
Potential countermeasures include: (i) proactively refreshing affected rows~\refreshBasedRowHammerDefenseCitations{}, (ii) relocating aggressor rows~\cite{saileshwar2022randomized,saxena2022aqua}, or (iii) throttling potentially harmful memory accesses~\cite{yaglikci2021blockhammer,canpolat2024breakhammer,greenfield2012throttling}.

\subsection{Motivation}
\label{motivation}

DRAM technology node scaling has increased the severity of \RH{} and RowPress~\rowHammerGetsWorseCitations{} and led to the appearance of a new read disturbance phenomenon, \CD{}~\cite{columndisturb}.
\revv{While \secref{sub:read_disturbance_mitigation} describes a wide range of techniques that mitigate \RHandRP{}, the same techniques are ineffective against \CD{} due to its fundamentally different nature.}
\revv{Specifically, these techniques rely on tracking activations and deploying their countermeasures at \emph{row granularity}~\mitigatingRowHammerAllCitations{}.}
\revv{However, as described in \secref{sub:dram_read_disturbance}, \CD{} induces bitflips across three consecutive subarrays (i.e., it occurs at \revv{subarray granularity}), affecting \emph{thousands of rows}.}
\revv{In this section, we discuss how naively mitigating \CD{} by (i) increasing the DRAM refresh rate, or (ii) performing simple modifications to the industry-standard \RH{} mitigation mechanism, PRAC~\pracCitations{}, incurs significant performance and energy overheads.}

\subsubsection{Mitigating \CD{} by Increasing DRAM Refresh Rate}
\label{sub:increasedref}

\begin{figure}[ht]
\centering
        \includegraphics[width=\columnwidth]{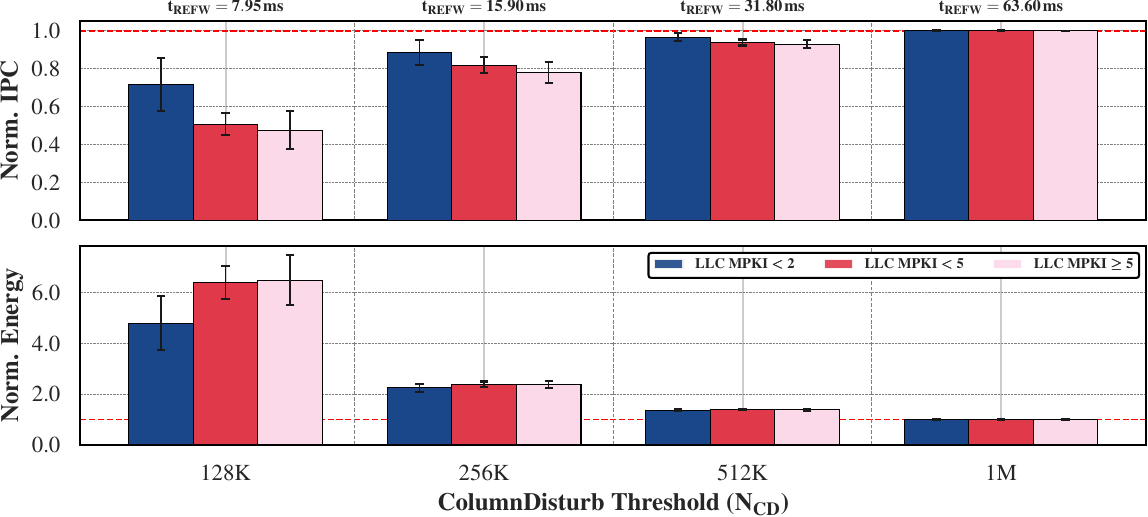}
        \caption{Normalized single-core performance and energy consumption of mitigating \CD{} by decreasing $\mathbf{t_{REFW}}$.}
        \label{fig:high-dram-refresh}
\end{figure}

The simplest method of mitigating \CD{} is to increase the refresh rate of DRAM by reducing $t_{REFW}$ and $t_{REFI}$.
This ensures that all affected DRAM cells are periodically refreshed before the \CD{} threshold (\nth), \revv{i.e., the number of row activations required to induce \CD{} bitflips,} is reached \revv{in any subarray}.
However, as both \CD{}~\cite{columndisturb} and our own analysis show, doing so significantly degrades system performance and energy efficiency for future DRAM chips.
Figure~\ref{fig:high-dram-refresh} shows the single-core performance (in instructions per cycle) and energy consumption of a system that mitigates \CD{} by increasing the DRAM refresh rate (reducing $t_{REFW}$), normalized to a baseline system with the default ($64\,\text{ms}$) refresh rate.
We observe that at the current \nth{} of \param{$1M$}, the reduced $t_{REFW}$ ($63.6\,\text{ms}$) degrades average system performance by just \param{0.03\%} and increases average DRAM energy consumption by \param{0.24\%}.
However, as \nth{} decreases, performance and energy consumption overheads caused by the lower $t_{REFW}$ increase significantly, particularly for workloads with high Last Level Cache (LLC) miss rates.
For $N_{CD}{=}128K$, the increased refresh rate degrades average system performance by over \param{50$\%$} and increases average DRAM energy consumption by \param{6$\times$}.
At this threshold, the required $t_{REFW}$ to protect against \CD{} is \param{$7.95\,\text{ms}$} (vs. \reviv{$64\,\text{ms}$} in DDR4).

\subsubsection{Mitigating \CD{} by Modifying PRAC}

Another potential way of mitigating \CD{} is by modifying the industry-standard \RH{} mitigation mechanism, PRAC~\pracCitations.
\revv{PRAC equips each DRAM row with a counter that tracks the number of activations to that row.
When a counter reaches a predefined threshold, PRAC raises an Alert Back-Off (ABO) signal, signaling the memory controller to issue an \texttt{RFM} command that preventively refreshes neighboring victim rows.}
To protect against \CD{}, one could: (i) reduce the ABO triggering threshold, and (ii) refresh \emph{all} rows across three subarrays when ABO is triggered.
Since an attacker can \revv{potentially} distribute a \CD{} attack across rows in a subarray to induce bitflips, ABO should be triggered after \emph{at most} $N_{CD}{/}S$ row activations, where $S$ is the subarray size.
For \reviv{$N_{CD}{=}1M$ and $S{=}1024$} (the subarray size reported in~\cite{columndisturb}), this threshold is \emph{only} \param{$1K$}.
As a result, after just $1K$ activations to a single row, this modified implementation would have to refresh $3K$ rows across 3 subarrays, incurring substantial performance and energy overheads.

\subsubsection{Our Goal \& Key Observation}
\label{motivation:double}

\begin{figure}[ht]
\centering
        \includegraphics[width=\columnwidth]{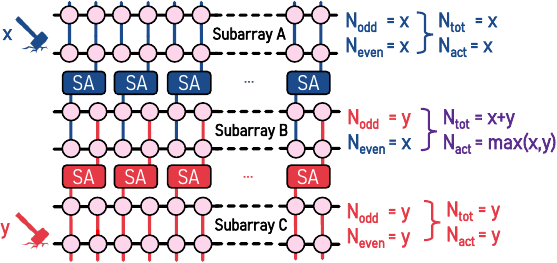}
        \caption{Example of ``\textit{double-counting}'' activations.}
        \label{fig:double-count-example}
\end{figure}

Our goal in \papername{} is to design the \emph{first} mechanisms that prevent \CD{} bitflips at low performance, energy, and area overheads.
We aim to achieve this by tracking activations at subarray granularity and issuing preventive refreshes to rows in \emph{all} affected subarrays.
Since each activation affects three consecutive subarrays, any mitigation mechanism must also account for activations in neighboring subarrays.
However, naively doing so may overestimate the activation count of the bitlines in a subarray for certain access patterns, leading to unnecessary refreshes and performance overheads.
Figure~\ref{fig:double-count-example} shows an example of such an access pattern across three DRAM subarrays.

An attacker performs $x$ ($y$) hammers in subarray A (C).
Due to the open-bitline architecture, subarray B shares its even (odd) bitlines with A (C).
As a result, both the odd and the even bitlines in A (C) \revv{experience} an \emph{identical} activation count of $x$ ($y$).
However, in subarray B, the activation count \revv{experienced by} the odd and even bitlines \emph{differs}, at $y$ and $x$, respectively.
A naive mechanism that counts the \emph{total} number ($N_{tot}$) of row activations affecting B reports an \revv{activation} count of $x{+}y$.
However, row activations in A (C) exclusively \revv{affect} the even (odd) bitlines in B.
Consequently, the actual hammer count of B ($N_{act}$) is the \emph{maximum} activation count of the odd and even bitlines (i.e., $\max(x,y)$).
When $x{=}y$, $N_{tot}{=}2{\times}N_{act}$.
We refer to this potential pitfall as ``\textit{double-counting}'' and take it into account when designing our deterministic mitigation mechanism, to reduce performance and energy overheads.

%% file: sections/03_mechanism.tex
\section{\papername{}}
\label{sec:mechanism}

\papername{} introduces two mechanisms that mitigate \CD{} by preventively refreshing DRAM rows:
(i)~\CGUARDD{}, a deterministic mitigation mechanism that tracks the number of times that the bitlines of a subarray have been hammered to issue \reviv{preventive} refreshes, and (ii)~\CGUARDP{}, a probabilistic mitigation mechanism that issues preventive refreshes with a predetermined probability.
To ensure no \CD{} bitflips occur, both mechanisms preventively refresh \emph{all} DRAM rows in a subarray and its \reviv{neighboring subarrays} before the \CD{} threshold ($N_{CD}$) is reached.
Due to the large number of rows that require refreshing, naively refreshing all potential victim rows at once incurs significant latency overheads by stalling accesses to the bank for up to $3K{\cdot}t_{RC}$ (i.e., $153\,\mu\text{s}$ for our configuration\revv{, see \secref{sec:methodology}}).
To avoid such latency overheads, both mechanisms distribute preventive refresh operations across time and perform them one row at a time by issuing \texttt{ACT+PRE} commands.
To mitigate \CD{}, \papername{} requires exposing the DRAM subarray mapping to the memory controller, similar to works that employ subarray-level parallelism (SALP)~\salpCitations{}.

\subsection{High-Level Architecture \& Shared Components}
\label{sub:key_components}
Both \CGUARDD{} (\CKD) and \CGUARDP{} (\CKP) consist of a \textit{trigger} mechanism and a \textit{countermeasure}.
The trigger mechanism \emph{decides when} to invoke the countermeasure for a subarray and/or its neighboring subarrays.
The countermeasure \emph{decides which} row to preventively refresh in an affected subarray.
\CKD{} and \CKP{} employ different trigger mechanisms (\secref{sub:cguardd} and~\secref{sub:cguardp}, respectively) but share the countermeasure, which is the \textbf{Row Pointer Table} (RPT).

\reviv{The role of the RPT} is to select \emph{which} row to preventively refresh when the trigger mechanism \textit{fires} for a subarray.
To do so, it stores a pointer to the next DRAM row to be preventively refreshed for each subarray.
We design the RPT as a table of counters, one for each of the $K$ subarrays of the system.
When the trigger mechanism \textit{fires} for a subarray $k$, \papername{} probes the corresponding RPT entry to identify which row to refresh in the targeted subarray.
The RPT first returns the counter's current value ($R_k$) for that subarray and then increments \revv{the counter} to point to the next row.
When the RPT is probed for the last row in a subarray, it is zeroed to point to the first row of the subarray.
Due to this round-robin selection, the row currently pointed to by the RPT is \revv{necessarily} the row that was \emph{least recently} preventively refreshed, and therefore has the highest hammer count within the subarray.

\subsection{\CGUARDD}
\label{sub:cguardd}

\CGUARDD{} (\CKD) aims to mitigate \CD{} with low performance and energy overheads.
To do so, it counts the number of activations affecting the bitlines of each DRAM subarray via activation counters and issues preventive refreshes to victim rows \emph{only when necessary} (i.e., each time \reviv{a counter reaches} a predefined threshold).
\CKD's trigger mechanism is aware of DRAM's open-bitline architecture and employs \revv{two} separate counters for the odd and even columns of each DRAM subarray to avoid ``\textit{double-counting}'' (\secref{motivation}) and reduce the number of \reviv{unnecessary} preventive refreshes issued.

\subsubsection{Design \& Operation} 

\begin{figure}[ht]
\centering   
        \includegraphics[width=\columnwidth]{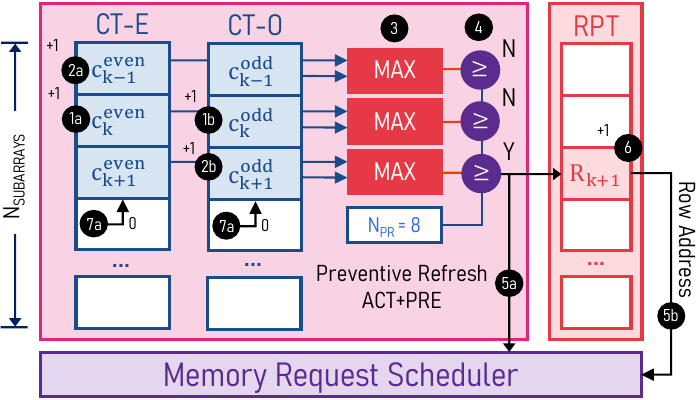}
        \caption{Overview of \papername{} in its \CKD{} configuration.}
        \label{fig:ColGuardD}
\end{figure}

Figure~\ref{fig:ColGuardD} shows the design of \CKD, consisting of a deterministic trigger mechanism and the countermeasure (RPT).
To accurately maintain the activation count of each subarray, \CKD{} introduces two counter tables that track the number of activations in the even and the odd columns of each subarray.
These are named \textbf{Counter Table-Even} (\CTE) and \textbf{Counter Table-Odd} (\CTO), respectively, and contain one entry per DRAM subarray in the system.
We refer to the number of activations in the even and the odd columns of a subarray as \cdcounter{k}{even} and \cdcounter{k}{odd}, respectively.

When the memory controller issues an \texttt{ACT} command to a row in subarray $k$, \CKD{} increments \cdcounter{k}{even} \circled{1a} and \cdcounter{k}{odd} \circled{1b}, because \emph{all} columns in $k$ are affected.
However, due to the open-bitline architecture, in the neighboring subarrays $k{-}1$ and $k{+}1$, \emph{only half} of the bitlines are connected to the sense amplifiers (\secref{sub:dram_org} and \secref{motivation}).
Thus, \CKD{} increments \cdcounter{k-1}{even} (i.e., the \CTE{} entry for $k{-}1$) \circled{2a} and \cdcounter{k+1}{odd} (i.e., the \CTO{} entry for $k{+}1$) \circled{2b}.\footnote{In the edge case \revv{where} $k$ is the first (last) subarray in the bank, \CKD{} \textit{only} performs these actions for $k{+}1$ ($k{-}1$).}
This way, \CKD{} avoids unnecessary counter increments for unaffected columns in neighboring subarrays.
To determine the activation count of an affected subarray \circled{3}, \CKD{} performs a \texttt{max} operation on the values of the \CTE{} and the \CTO{}, and calculates the maximum activation count \revv{observed} across their columns.
It then compares the result to a predetermined \textit{preventive refresh threshold} ($N_{PR}$) \circled{4}, \revv{which we set to $N_{CD}{/}S$, where $S$ is the number of rows in a subarray.}%
\footnote{\revv{Our security analysis in \secref{security:d} shows that setting $N_{PR}$ to this value prevents \CD{} bitflips under a worst-case access pattern.}}\textsuperscript{,}%
\footnote{\revv{For reasons explained in \secref{security:activations}, we always subtract $2S$ from $N_{CD}$ before calculating $N_{PR}$ to account for activations caused by \texttt{REF} commands.}}
If the result of this comparison is \textit{True} (as in this case for subarray $k{+}1$), \CKD{} issues a preventive refresh to the affected subarray \circled{5a}, by retrieving the next row to be refreshed via the corresponding \RPT{} entry \circled{5b} and then incrementing \revv{the entry} to point to the next row of the subarray \circled{6}.
After issuing the refresh, \CKD{} \reviv{resets} the \CTE{} and \CTO{} entries for the subarray that triggered the refresh \circled{7a} and \circled{7b}.

\subsection{\CGUARDP} 
\label{sub:cguardp}

\CGUARDP{} (\CKP) aims to mitigate \CD{} bitflips with low hardware complexity.
To achieve this, \CKP{} uses a probabilistic trigger mechanism that requires no activation counters (i.e., it is stateless) and is configurable to provide a desired level of \reviv{protection against \CD{} bitflips}.

\subsubsection{Design \& Operation}
\reviv{When} the memory controller issues an \texttt{ACT} command to a row in subarray $k$, \CKP{} ``\textit{flips a coin}'' with a \textit{preventive refresh probability} ($P_{PR}$) to decide whether \revv{or not} to perform a \revv{preventive refresh}.
\revv{We calculate $P_{PR}$ so that the probability of a \CD{} bitflip (i.e., failing to refresh all rows in a subarray before the \CD{} threshold is reached) remains under a configurable upper bound.}\footnote{\revv{A similar probabilistic mitigation mechanism, PARA, was proposed by Kim et al.~\cite{kim2014flipping} to mitigate \RH{} bitflips.
Our security analysis in \secref{security:p} uses the detailed security analysis of PARA in HiRA~\cite{yaglikci2022hira} to calculate $P_{PR}$.}}
If the coin flip \reviv{``\textit{results in a hit}''}, \CKP{} issues three \reviv{mitigation requests}, one for $k$ and one for each of its neighbors $k{-}1$ and $k{+}1$.
To identify which rows to refresh within the affected subarrays, similar to \CKD{}, \CKP{} probes the RPT for all three subarrays, and retrieves $R_{k-1}$, $R_k$, and $R_{k+1}$.
After issuing the preventive refreshes, \CKP{} increments the corresponding RPT entries to point to the next row.

%% file: sections/04_security.tex
\section{Security Analysis}

\noindent\textbf{Worst-case access pattern:}
To overwhelm \papername, the worst-case access pattern aims to trigger as many refresh operations as possible within its hammering budget, defined by DRAM timing constraints. 
To achieve this, the worst-case access pattern leverages the fact that hammering a row disturbs \emph{all} other rows in the same subarray and \revv{all rows} in adjacent subarrays. 
As such, hammering \emph{any} row in a single subarray causes the highest number of preventive refresh operations.

\subsection{Security Analysis of \CGUARDD}
\label{security:d}

We use ``\textit{proof by induction}'' to prove that \CGUARDD{} (\CKD) prevents \CD{} bitflips in three steps.

First, we calculate the hammer count ($HC$) that the odd and even cells of a row (i.e., the cells residing in the odd and even columns, respectively) are exposed to as the sum of the row activation count ($n_k$) and the refresh count ($r_k$) targeting the row's subarray $k$ and the neighboring subarray $k^*$ ($n_{k^*}$ and $r_{k^*}$), where $k^*$ is $k{-}1$ or $k{+}1$. 
Therefore, $HC_{odd} = \Sigma \{n_{k}, r_{k}, n_{k-1}, r_{k-1}\}$ and $HC_{even} = \Sigma \{n_{k}, r_{k}, n_{k+1}, r_{k+1}\}$, respectively.
Second, we define the \emph{base case} and the \emph{induction hypothesis}.
Third, we prove the \emph{induction hypothesis}.

\noindent\textbf{Base Case:} 
DRAM is powered on.
$HC_{odd}{=}0$ and $HC_{even}{=}0$ for all cells.
Therefore, all cells are secure.

\noindent\textbf{Induction Hypothesis:}
A randomly-chosen row $i$ in subarray $k$ has just been refreshed by activation $n$ (i.e., $HC_{odd}(i,n){=}0$ and $HC_{even}(i,n){=}0$), and is thus secure.
\emph{We hypothesize} that row $i$ will \emph{remain} secure, meaning that \emph{either}
(i) not enough activations to cause a \CD{} bitflip in row $i$ will occur, \emph{or}
(ii) row $i$ will have already been preventively refreshed again \emph{before} the sum of any combination of row activations and refreshes in subarray $k$ and its neighboring subarrays ($k{-}1$ and $k{+}1$) reaches $N_{CD}$.

\noindent\textbf{Induction Proof:} 
From the \textit{induction hypothesis}, $HC_{odd}(i,n){=}0$, $HC_{even}(i,n){=}0$.
Let $n'{=}n{+}N_{CD}$. 
At activation $n'$, row $i$'s hammer counts for odd and even columns ($HC_{odd}(i, n')$ and $HC_{even}(i, n')$) become the sum of all row activations and refreshes that happened in subarray $k$ and its two adjacent subarrays ($\Sigma \{n_{k}, r_{k}, n_{k-1}, r_{k-1}\}$ and $\Sigma \{n_{k}, r_{k}, n_{k+1}, r_{k+1}\}$), respectively. 
As described in \secref{sub:cguardd}, \CKD{} resets \emph{both} the \CTO{} and \CTE{} entries for subarray $k$ whenever either of them reaches $N_{PR}$, and a preventive refresh in subarray $k$ fires at every such reset.
Between any two consecutive resets, \emph{neither} entry can climb past $N_{PR}$. 
Otherwise, the preventive refresh would have been issued earlier. 
Let $r_k$ be the number of preventive refreshes (and thus also resets) that have been issued to subarray $k$ between activation $n$ and $n'$.
Hence $HC_{odd}(i,n') {\leq} r_k {\cdot} N_{PR}$, and $HC_{even}(i,n') {\leq} r_k {\cdot} N_{PR}$, leading to Inequality~\ref{exp:rk-evaluation}.

\vspace{-0.75em}
\begin{equation}
\label{exp:rk-evaluation}
\max(HC_{odd}(i,n'), HC_{even}(i,n')) \leq r_k \cdot N_{PR}
\end{equation}

\noindent Based on the values of $HC_{odd}(i,n')$, $HC_{even}(i,n')$ and $N_{CD}$, we consider the following two cases.

\noindent\textit{\underline{Case 1:}} The trivial case where the hammer count is \emph{too low} to induce bitflips (i.e., $\max(HC_{odd}(i,n'), HC_{even}(i,n')) {<} N_{CD}$). 

\noindent\textit{\underline{Case 2:}} The hammer count is \emph{high enough} to induce bitflips (i.e., $\max(HC_{odd}(i,n'), HC_{even}(i,n')) {\geq} N_{CD}$).
By transitivity from the definition of \textit{Case 2} and Inequality~\ref{exp:rk-evaluation}, we derive the following inequality:

\vspace{-0.75em}
\begin{equation}
\label{exp:secure_d-second_condition}
N_{CD} \leq \max(HC_{odd}(i,n'), HC_{even}(i,n')) \leq r_k {\cdot} N_{PR} 
\end{equation}

\noindent As defined in \secref{sub:cguardd}, $N_{PR}{=}N_{CD}{/}S$.
By substituting $N_{PR}$ and simplifying $N_{CD}$ in Inequality~\ref{exp:secure_d-second_condition}, we derive $r_k \geq S$.
To remain secure, row $i$ must be preventively refreshed before or by activation $n'$.
Since row $i$ is the \emph{most recent} row to be preventively refreshed (in subarray $k$), it will be the \emph{last} row to \emph{again} be preventively refreshed (in subarray $k$), as the \RPT{} selects rows in a round-robin manner.
For row $i$ to be refreshed again, the number of preventive refreshes issued to subarray $k$ ($r_k$) must be \emph{at least} $S$, where $S$ is the number of rows in a subarray (i.e., $r_k \geq S$).
However, this was already derived above, satisfying the condition for \textit{Case 2} remaining secure, and thus proving the \textit{induction hypothesis}.

\subsection{Security Analysis of \CGUARDP}
\label{security:p}

\CGUARDP{} (\CKP) provides a configurable level of security by adjusting the \textit{preventive refresh probability} ($P_{PR}$).
To configure $P_{PR}$, we follow the methodology proposed in HiRA~\cite{yaglikci2022hira} for \RH{} \revv{and PARA~\cite{kim2014flipping}}.
\revv{Specifically, we consider the worst-case scenario consisting of the \emph{maximum} number of \CD{} attacks (i.e., attempts to induce \CD{} bitflips) possible within a refresh window.} 
Let: 

\begin{itemize}
    \item $HC(r)$ be the number of hammers a row $r$ has experienced since \revv{it was} last preventively refreshed.
        As \CKP{} is not aware of the open-bitline architecture, $HC$ does not distinguish between odd and even bitlines.
    \item $X_i$ be a random variable that models whether \CKP{} issues a preventive refresh to a subarray following a hammer (i.e., any kind of activation) $i$ to the subarray \emph{or} its neighbors.
    \item $M_N$ be a random variable that models the number of preventive refreshes issued to a subarray after $N$ activations to it or its neighbors.
    \item $A$ be the \emph{maximum} number of attacks that can occur within a refresh window.
    \item $t_F$ be the shortest time in which an attack can fail.
    \item $P_1$ be the probability of a \emph{single} successful attack.
    \item $P_{REFW}$ ($P_Y$) be the probability of \emph{at least} one successful attack during a refresh window (a year).
\end{itemize}

\noindent We will show that \emph{any} row in \emph{any} subarray starting from an initial secure state $HC(r){=}0$ (i.e., just initialized or preventively refreshed), will suffer a successful attack after $N_{CD}$ hammers with a probability of $P_1$. 
We then extend this probability bound to an entire refresh window ($P_{REFW}$), and an entire year ($P_Y$).

\subsubsection{Cumulative Distribution Function (CDF) of $\mathbf{M_N}$} \CKP{} randomly issues a preventive refresh following each \texttt{ACT} to a subarray or its neighbors with a probability $P_{PR}$.
This occurs via a ``\textit{coin-flip}'', independent of any other event (i.e., a Bernoulli trial). 
$M_N$ counts the number of preventive refreshes issued to the \textit{middle} subarray after $N$ activations to it or its neighbors.
As a result: $M_N{=}\sum_{i=1}^{N}X_i$.
The sum of $N$ independent Bernoulli trials with probability $P_{PR}$ follows a \textit{Binomial distribution} $B(N,P_{PR})$.\footnote{The Binomial distribution is derived from ``\textit{the number of $x$ occurrences in $n$ independent trials}''~\cite{bincdf-beta} (page 48).}
The CDF of the binomial distribution is the \textit{regularized incomplete beta function}~\cite{bincdf-beta} ($I_q$), where $q{=}1{-}P_{PR}$, therefore: 

\vspace{-0.75em}
\begin{equation}
\label{eq:cdf}
P(M_N \le k) = I_{q}(N - k, 1 + k)
\end{equation}

\subsubsection{Probability of a single successful attack}
A single attack succeeds (with probability $P_1$) when \CGUARDP{} issues \emph{fewer} than $S$ preventive refreshes to a subarray following $N_{CD}{-}1$ activations to it or its neighbors.
Eq.~\ref{eq:single_attack} formulates $P_1$ by substituting $N$ with $N_{CD}{-}1$ and $k$ with $S{-}1$ in Eq.~\ref{eq:cdf}.

\vspace{-0.75em}
\begin{equation}
\label{eq:single_attack}
P_1 = I_{q}(N_{CD} - S, S)
\end{equation}

\subsubsection{Maximum number of attacks in a refresh window}
An attacker can mount multiple attacks within a refresh window ($t_{REFW}$) of which only one \emph{needs} to succeed.
To evaluate the worst-case scenario, we calculate the \textit{maximum} number of attacks ($A$) that the attacker can mount.
Following the process described in HiRA~\cite{yaglikci2022hira}, we calculate $A$ by assuming $A{-}1$ attacks that take (fail in) the shortest possible duration ($t_F$), followed by a single successful attack.
A single attack fails when all rows within a subarray are preventively refreshed.
The attacker \textit{initiates} the attack with a single row activation.
To stop an attack, \CKP{} has to issue at least $S$ mitigation requests, each of which refreshes one row in three consecutive subarrays.
Therefore, $t_F{=}(1{+}3S){\cdot}t_{RC}$, and $A{=}t_{REFW}{/}t_F{+}1$.

\subsubsection{Probability of a successful attack within a refresh window}
The probability of \textit{at least} one successful attack within a refresh window ($P_{REFW}$) is the complement of the probability of all $A$ attacks failing.
Since the outcomes of the $A$ attacks are mutually independent, $P(\text{all fail}) = P(\text{one fails})^A$.
In turn, $P(\text{one fails})$ is the complement of the probability of a single successful attack ($P_1$). As a result:

\vspace{-0.75em}
\begin{equation}
\label{eq:refw}
P_{REFW}=1-(1-P_1)^{A}
\end{equation}

\subsubsection{Probability of a successful attack in a year}
A year consists of $492.7{\times}10^6$ DDR4 refresh windows. 
The probability of \emph{at least} one successful attack within a year ($P_Y$) is the complement of no attacks succeeding within those refresh windows. 
Following what was described for $P_{REFW}$:

\vspace{-0.75em}
\begin{equation}
\label{eq:year}
P_{Y} = 1 - (1-P_{REFW})^{492.7{\times}10^{6}}
\end{equation}

\subsection{Configuration of \CGUARDP}
In Table~\ref{tab:parav2}\revv{,} we first set target $P_{Y}$ values and calculate $P_{REFW}$ (Eq.~\ref{eq:year}) and $P_1$ (Eq.~\ref{eq:refw}).
We then solve Eq.~\ref{eq:single_attack} to calculate $P_{PR}$ for three different \CD{} thresholds (\nth) and for a subarray size of $1K$ rows (as shown in~\cite{columndisturb}). 
We make two observations. First, providing the same security \revv{level} for lower thresholds requires significantly higher $P_{PR}$.
For example, for $P_Y${=}$10^{-6}$, when \nth{} drops from $1M$ to $16K$, $P_{PR}$ increases by $72\times$. 
Second, for the same \nth, even slight increases to $P_{PR}$ provide significantly higher security.
For example, with $N_{CD}{=}128K$ and $P_{PR}{=}10^{-2}$, $P_Y{=}10^{-3}$.
However, when $P_{PR}$ slightly increases to $1.07{\times}10^{-2}$, $P_Y$ drops to $10^{-12}$.
For the rest of this paper, $P_{PR}$ is always calculated for $P_Y{=}10^{-12}$ unless stated otherwise.

\input{tables/PARAV2-Analysis}

\subsubsection{\CGUARDP{} Monte Carlo Security Analysis}
\label{sec:montecarlo}

\begin{figure}[ht]
\centering   
        \includegraphics[width=\columnwidth]{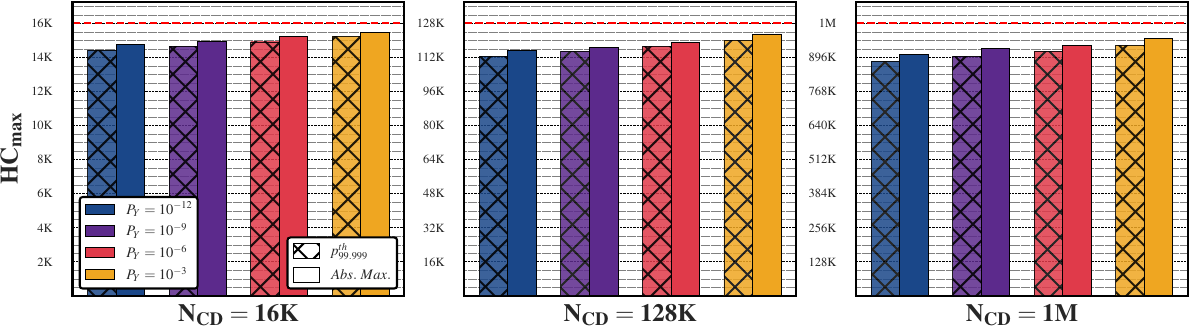}
        \caption{$p_{99.999}^{th}$ and absolute maximum value of $HC_{max}$ for the different configurations of Table~\ref{tab:parav2}.}
        \label{fig:montecarlo}
\end{figure}

We demonstrate \CGUARDP's security by performing a Monte Carlo simulation of a DRAM bank under continuous \CD{} attack.
\revv{Specifically, we simulate continuously hammering a randomly selected row in a randomly selected subarray.
We then measure the \emph{maximum} number of times ($HC_{max}$) \revv{the cells residing in the odd or even columns of \emph{any} row} were hammered without \CKP{} preventively refreshing the row during an entire refresh window, and repeat this experiment $1M$ times.
Figure~\ref{fig:montecarlo} shows the ${99.999}_{th}$ percentile ($p_{99.999}^{th}$) (hatched bars) and overall maximum value (solid bars) of $HC_{max}$ for the different configurations of Table~\ref{tab:parav2}.}

\noindent We make two key observations.
First, across all configurations, $HC_{max}$ remains well below \revv{$N_{CD}$} (represented by the red line).
Second, as the security guarantees tighten (i.e., $P_Y$ decreases), the gap between the absolute maximum value of $HC_{max}$ and $N_{CD}$ increases.
For example, when $P_Y$ reduces from $10^{-3}$ to $10^{-12}$ at $N_{CD}$ equal to $128K$, this gap increases from $5.5K$ to $13K$ activations, representing a ``\textit{safety margin}'' of $4.2\%$ and $9.9\%$, relative to $N_{CD}$.
\revv{We attribute this to the fact that achieving lower values of $P_Y$ for the same $N_{CD}$ requires setting higher values of $P_{PR}$.\footnote{By backtracking through Equations~\ref{eq:year}, \ref{eq:refw} and \ref{eq:single_attack}, we can derive that $P_{Y}$ is monotonically decreasing with $P_{PR}$. Table~\ref{tab:parav2} demonstrates this relationship.}}
\revv{A higher $P_{PR}$ in turn leads to \CKP{} being more likely to issue preventive refreshes, reducing $HC_{max}$.}

\subsection{Counting Activations in Subarrays}
\label{security:activations}

Both \CKD{} and \CKP{} rely on tracking activations in all subarrays. 
Row activations occur: (i) \textit{explicitly} via \texttt{ACT} commands or (ii) \textit{implicitly} via periodic refresh commands (\texttt{REF}) and \RH{} preventive refreshes (e.g., \texttt{RFM} commands in DDR5).
While \papername{} maps \texttt{ACT}s to subarrays, \textit{implicit} activations \revv{\emph{do not}} specify a row address, \revv{and therefore cannot be directly mapped to a subarray at the level of the memory controller, where \papername{} resides.}
To remain secure, \papername{} handles \textit{implicit} activations as \revv{described below}.

\subsubsection{Periodic Refreshes}
While \texttt{REF} commands do not specify a DRAM row or subarray, \emph{all} DRAM rows are refreshed \emph{exactly} once within a refresh window.
During the window, the last \revv{row} to be refreshed will experience $S$ hammers from refreshes in a neighboring subarray and $S{-}1$ hammers from refreshing \revv{rows} in the same subarray for a total of $2S{-}1$ hammers.
Taking this into account, we \emph{always} calculate $N_{PR}$ after subtracting $2S$ from the actual value of $N_{CD}$. 
This acts as a safety margin and ensures that preventive refreshes will not inadvertently result in \CD{} bitflips. 
To reduce performance overheads, DRAM standards~\cite{jedec2012jedec,jedec2020ddr5} allow postponing \texttt{REF} commands up to \param{8} (\param{4}) times for DDR4 (DDR5), performing back-to-back refresh operations after postponement.
To account for \texttt{REF} commands that might be postponed to the next refresh window, we further adjust $N_{CD}$ by subtracting an additional $4\times Q$ for DDR5 ($8\times Q$ for DDR4), where $Q$ is the number of rows refreshed per \texttt{REF} command.\footnote{This reduction is negligible for the $N_{CD}$ values that are evaluated in~\secref{sec:eval}.} 

\subsubsection{Preventive Refreshes}
When combining \papername{} with a \RH{} mitigation mechanism, the memory controller issues preventive refreshes for both mechanisms.
\papername{} treats its own \texttt{ACT} commands as regular \texttt{ACTs}, i.e., \CKD{} increments the appropriate counters, and \CKP{} ``\textit{flips its coin}'' for the mitigation requests it issues. 
\RH{} mitigation mechanisms refresh rows by: (i) issuing \texttt{ACT+PRE} commands~\cite{olgun2024abacus,kim2014flipping} (handled \revv{as explicit activations}), (ii) issuing extra \texttt{REF} commands~\cite{bostanci2024comet}, or (iii) issuing \texttt{RFM} commands in DDR5~\cite{jedecddr5c,canpolat2024understanding,canpolat2025chronus,woo2025qprac}. 
To handle \texttt{RFM} or extra \texttt{REF} commands, \CKD{} increments \emph{all} \CTO{} and \CTE{} entries in the affected banks, since a \texttt{REF} or \texttt{RFM} command may affect any subarray.
Similarly, \CKP{} ``\textit{flips its coin}'' for all subarrays.
Thus, \papername{} can protect against \CD{} bitflips from \RH{} preventive refreshes.
Since \RH{} mitigations also track \papername's \texttt{ACT+PRE} commands, the inverse is also true.

%% file: tables/PARAV2-Analysis.tex
\newcolumntype{Y}{>{\centering\arraybackslash}X}

\begin{table}[ht]
\caption{Configuration parameters of \CGUARDP{}.}
\label{tab:parav2}
\renewcommand{\arraystretch}{1.2}
\centering
\footnotesize %
\begin{tabularx}{1.02\columnwidth}{|Y|Y|Y|Y|Y|}
\hline
\boldmath{$P_{Y}$} & $10^{-3}$ & $10^{-6}$ & $10^{-9}$ & $10^{-12}$ \\ \hline \hline 
\boldmath{$P_{REFW}$} & {\footnotesize $2.03{\times} 10^{-12}$} & {\footnotesize $2.03{\times} 10^{-15}$} & {\footnotesize $2.03{\times} 10^{-18}$} & {\footnotesize $2.03{\times} 10^{-21}$} \\ \hline \hline
\boldmath{$P_1$} & {\footnotesize $4.39{\times} 10^{-15}$} & {\footnotesize $4.38{\times} 10^{-18}$} & {\footnotesize $4.38{\times} 10^{-21}$} & {\footnotesize $4.38{\times} 10^{-24}$} \\ \hline \hline
\boldmath{$N_{CD}$} & \multicolumn{4}{c|}{\textbf{Preventive Refresh Probability \bfseries\boldmath($P_{PR}$)}} \\ \hline
$1M$ & {\footnotesize $1.23{\times} 10^{-3}$} & {\footnotesize $1.26{\times} 10^{-3}$} & {\footnotesize $1.29{\times} 10^{-3}$} & {\footnotesize $1.32{\times} 10^{-3}$} \\ \hline
$128K$ & {\footnotesize $1.00{\times} 10^{-2}$} & {\footnotesize $1.02{\times} 10^{-2}$} & {\footnotesize $1.05{\times} 10^{-2}$} & {\footnotesize $1.07{\times} 10^{-2}$} \\ \hline
$16K$ & {\footnotesize $8.92{\times} 10^{-2}$} & {\footnotesize $9.13{\times} 10^{-2}$} & {\footnotesize $9.32 {\times} 10^{-2}$} & {\footnotesize $9.50{\times} 10^{-2}$} \\ \hline
\end{tabularx}
\end{table}

%% file: sections/05_methodology.tex
\section{Methodology}
\label{sec:methodology}

We evaluate the impact of \papername{} on system performance and energy consumption using Ramulator 2.0~\ramulatorCitations, a cycle-accurate DRAM simulator, and DRAMPower~\cite{drampower,steiner2025drampower}.
Table~\ref{configs} summarizes the configuration of our simulated system.

\input{tables/evaluated-system-configuration}

\noindent\textbf{Workloads.} 
We use 62 single-core workloads from 5 benchmark suites: SPEC CPU2006~\cite{spec2k6}, SPEC CPU2017~\cite{spec2017}, TPC~\cite{tpcweb}, MediaBench~\cite{fritts2k9media}, and YCSB~\cite{ycsb}.
To simulate multi-core workloads we first organize the workloads into 3 categories based on their row buffer misses per kilo-instruction (RBMPKI): (i) Low, (ii) Medium, and (iii) High (see Table~\ref{table:workloads}).
Second, by randomly selecting traces from the aforementioned categories we create 10 four-core workload mixes for each one of the following categories: LLLL, MMMM, HHHH, LLMM, MMHH, LLHH for a total of 60 workload mixes.
We simulate single- and multi-core workloads until each evaluated workload (and each core) executes at least 100M instructions. 

\input{tables/rbmpki-characterization}

\noindent\textbf{Evaluated Systems.} 
We evaluate:
\textbf{(1) \CGUARDD{}} (\CKD): The mitigation mechanism of~\secref{sub:cguardd}, 
\textbf{(2) \CGUARDP3} (\CKP[3]): The mitigation mechanism of~\secref{sub:cguardp} configured for $P_Y$ of $10^{-3}$, 
\textbf{(3) \CGUARDP12} (\CKP[12]): The same mechanism for $P_Y{=}10^{-12}$,
\textbf{(4) \papername-S} (\CKS): an alternative design of \CKD{} that uses a single counter per subarray and does not account for ``\textit{double-counting}'',
\textbf{(5) SALP+\CGUARDD{}} (\CKDSALP): a modified version of \CGUARDD{} that employs subarray-level parallelism~\cite{kim2012case} (SALP),\footnote{\revv{SALP-enabled systems use the MASA~\cite{kim2012case}} SALP configuration.}
and \textbf{(6) SALP+\CGUARDP{}} (\CKPSALP): a modified version of \CKP[12] that employs SALP. 
For the evaluation of \papername{} with \RH{} mitigation mechanisms, we integrate \papername{} with Graphene~\cite{park2020graphene}, PRAC~\pracCitations, and Hydra~\cite{qureshi2022hydra} configured for a \RH{} threshold ($N_{RH}$) of 128.\footnote{PRAC~\pracCitations{} was introduced in the JEDEC DDR5 specification~\cite{jedecddr5c}. For fairness, all configurations that employ PRAC use DDR5.}

%% file: tables/evaluated-system-configuration.tex
\begin{table}[h]
\centering
\caption{Simulated System Configuration.}
\label{configs}
\resizebox{\linewidth}{!}{
\begin{tabular}{ll}
\hline
\textbf{Processor}                                                   & \begin{tabular}[c]{@{}l@{}} 1 or 4 cores, 3.6 GHz clock frequency,4-wide issue, 128-entry instr. window\end{tabular}  \\ \hline
\textbf{Page Allocator}                                       & \begin{tabular}[c]{@{}l@{}} 4KB Buddy Allocator~\cite{buddy}, pre-fragmented to 35\% at the 2MB granularity.\end{tabular}  \\ \hline
\textbf{DRAM}                                                        & \begin{tabular}[c]{@{}l@{}} 16GB DDR4, 1 channel, 2 rank/channel, 4 bank groups,\\ 4 banks/bank group, 64 subarrays/bank, 1K rows/subarray, 3200 MT/s\end{tabular}  \\ \hline
\begin{tabular}[c]{@{}l@{}}\textbf{Memory Ctrl.}\end{tabular} & \begin{tabular}[c]{@{}l@{}}64-entry read/write requests queue,\\Scheduling policy: FR-FCFS~\cite{rixner2000memory,zuravleff1997controller} with open-row policy~\cite{kaseridis2011minimalistic,moscibroda2007memory},\\Address mapping: RoBaRaCoCh~\cite{ramulator2github,canpolat2025chronus}\\
\end{tabular}   \\ \hline
\textbf{Last-Level Cache}& \begin{tabular}[c]{@{}l@{}} 2 MiB per core \end{tabular}  \\ \hline
\end{tabular}}
\end{table}

%% file: tables/rbmpki-characterization.tex
\begin{table}[ht]
\caption{Single-core Workloads Organized by RBMPKI.}
\label{table:workloads}
\centering
\setlength{\tabcolsep}{3pt}
\renewcommand{\arraystretch}{0.75}
\renewcommand{\tabularxcolumn}[1]{>{\scriptsize\arraybackslash}m{#1}}
\begin{tabularx}{\linewidth}{|>{\scriptsize}c||X|}
\hline
\multicolumn{1}{|c||}{\scriptsize\textbf{RBMPKI}} & \multicolumn{1}{c|}{\scriptsize\textbf{Workloads}} \\ \hline
\begin{tabular}[c]{@{}c@{}}$[10+)$\\(High)\end{tabular} & 519.lbm, 459.GemsFDTD, 450.soplex, h264\_decode, 520.omnetpp, 433.milc, 434.zeusmp, bfs\_dblp, 429.mcf, 549.fotonik3d, 470.lbm, bfs\_ny, bfs\_cm2003, 437.leslie3d, gups \\ \hline
\begin{tabular}[c]{@{}c@{}}$[2, 10)$\\(Med.)\end{tabular} & 510.parest, 462.libquantum, tpch2, wc\_8443, ycsb\_aserver, 473.astar, jp2\_decode, 436.cactusADM, 557.xz, ycsb\_cserver, ycsb\_eserver, 471.omnetpp, 483.xalancbmk, 505.mcf, wc\_map0, jp2\_encode, tpch17, ycsb\_bserver, tpcc64, 482.sphinx3 \\ \hline
\begin{tabular}[c]{@{}c@{}}$[0, 2)$\\(Low)\end{tabular} & 502.gcc, 544.nab, h264\_encode, 507.cactuBSSN, 525.x264, ycsb\_dserver, 531.deepsjeng, 526.blender, 435.gromacs, 523.xalanc\-bmk, 447.dealII, 508.namd, 538.imagick, 445.gobmk, 444.namd, 464.h264ref, ycsb\_abgsave, 458.sjeng, 541.leela, tpch6, 511.povray, 456.hmmer, 481.wrf, grep\_map0, 500.perlbench, 403.gcc, 401.bzip2 \\ \hline
\end{tabularx}
\end{table}

%% file: sections/06_evaluation.tex
\begin{figure*}[th]
\centering  
        \includegraphics[width=\textwidth]{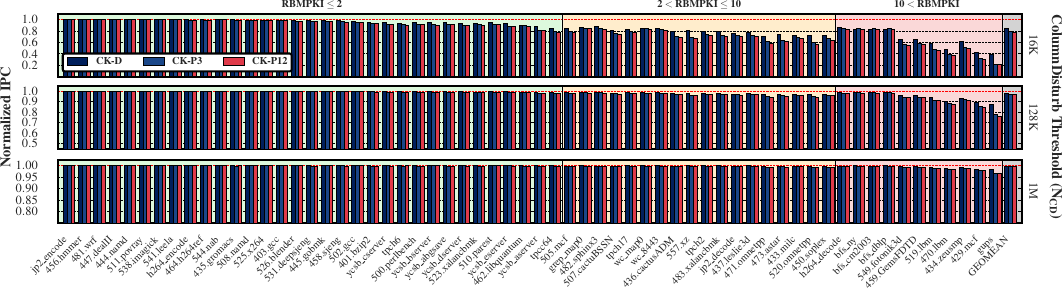}
        \caption{\revv{Performance impact of \papername{} on single-core workloads for three different \CD{} thresholds.}}
        \label{fig:detailed_single_core_ipc}
\end{figure*}

\section{Evaluation}
\label{sec:eval}

We evaluate:
1) \papername's impact on system performance and DRAM energy consumption,
2) the performance impact of \papername{} under adversarial workloads,
3) the impact of combining \papername{} with different \RH{} mitigation mechanisms \revv{(i.e., Graphene~\cite{park2020graphene}, PRAC~\pracCitations{}, and Hydra~\cite{qureshi2022hydra})},
4) the impact of different DRAM subarray sizes on \papername's performance,
5) the impact of memory fragmentation and different page allocators~\cite{buddy,hsieh2016transparent} on \papername's performance, and
6) the performance of \papername{} in systems that employ subarray-level parallelism (SALP)~\salpCitations{}.

\subsection{Single-Core Evaluation}
\label{eval:single}

\subsubsection{Performance} 
Figure~\ref{fig:detailed_single_core_ipc} presents \papername's performance impact on single-core workloads.
The top, middle, and bottom subplots show results for $N_{CD}{=}16K$, $128K$, and $1M$, respectively.
The x-axis depicts different workloads, organized from left to right by increasing RBMPKI.
Different background colors denote different RBMPKI categories.
The y-axis shows the instructions-per-cycle (IPC) of each workload normalized to a baseline with no \CD{} mitigation.

We draw three major conclusions. 
First, for single-core workloads, at the current threshold ($1M$), \CKD{}, \CKP[3], and \CKP[12] incur \revv{small} performance degradation with minimum normalized IPCs of \param{\IpcSCMinDHi}, \param{\IpcSCMinPaHi}, and \param{\IpcSCMinPbHi}, respectively.
Second, for $N_{CD}{=}128K$, all three variants retain good average performance with a geomean normalized IPC of \param{\IpcSCGmDMi}, \param{\IpcSCGmPaMi}, and \param{\IpcSCGmPbMi}, respectively.
However, in high-RBMPKI workloads, we observe that \CKP[3] (\CKP[12]) shows non-negligible performance degradation.
For example, the normalized IPC for \texttt{gups} is \param{\IpcSCGupsPaMi} (\param{\IpcSCGupsPbMi}).
Third, for $N_{CD}{=}16K$, all mitigations incur significant performance degradation.
Specifically, \CKD{} \revv{has} a geomean normalized IPC of \param{\IpcSCGmDLo}, while \CKP[3] (\CKP[12]) has a geomean normalized IPC of \param{\IpcSCGmPaLo} (\param{\IpcSCGmPbLo}).
To provide security at this very low threshold, \CKP[3] (\CKP[12]) issues on average \param{\MrSCRatioPaOverDLo$\times$} (\param{\MrSCRatioPbOverDLo$\times$}) more \revv{preventive refreshes} than \CKD{}.

\subsubsection{Energy Consumption}
\begin{figure}[ht]
\centering   
        \includegraphics[width=\columnwidth]{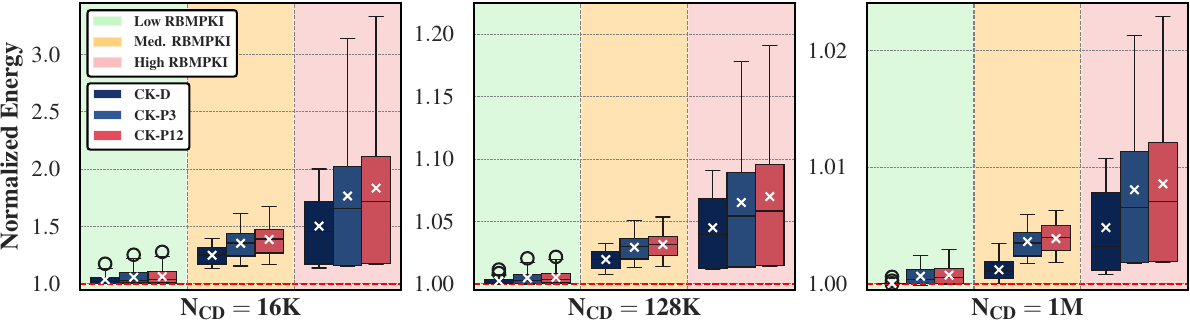}
        \caption{\revv{Effect of \papername{} on DRAM energy consumption for single-core workloads.}}
        \label{fig:single_core_energy}
\end{figure}

Figure~\ref{fig:single_core_energy} presents \papername's impact on energy consumption for single-core workloads, with results grouped by RBMPKI.
The left, middle, and right subplots show results for $N_{CD}{=}16K$, $128K$, and $1M$, respectively.
Each boxplot corresponds to a particular \papername{} variant and RBMPKI category (denoted by background color), and shows the distribution of energy consumption normalized to a baseline with no \CD{} mitigation.

We draw three major conclusions.
First, at $N_{CD}{=}1M$, the average energy increase remains \revv{small} (${<}2\%$) for all \papername{} variants.
Second, at $128K$, \CKD{} incurs a small increase, with an average (maximum) increase of \param{\EnSCAvgDMi\%} (\param{\EnSCMaxDMi\%}), while \CKP[3] and \CKP[12] incur average (maximum) increases of \param{\EnSCAvgPaMi\%} (\param{\EnSCMaxPaMi\%}) and \param{\EnSCAvgPbMi\%} (\param{\EnSCMaxPbMi\%}), respectively.
Third, at $16K$, all variants significantly increase energy consumption by \param{\EnSCAvgDLo\%} for \CKD{}, \param{\EnSCAvgPaLo\%} for \CKP[3], and \param{\EnSCAvgPbLo\%} for \CKP[12], on average.
At this threshold, we observe that due to \CKD{} not issuing any ``\textit{unnecessary}'' refreshes, for the high RBMPKI category, it keeps the maximum energy increase to under \param{\EnSCHighMaxDLo$\times$}, compared to \param{\EnSCHighMaxPaLo$\times$} for \CKP[3] and \param{\EnSCHighMaxPbLo$\times$} for \CKP[12].

\subsection{Multi-Core Evaluation}
\label{eval:multi}

\subsubsection{Performance}
\begin{figure}[ht]
\centering
    \includegraphics[width=\columnwidth]{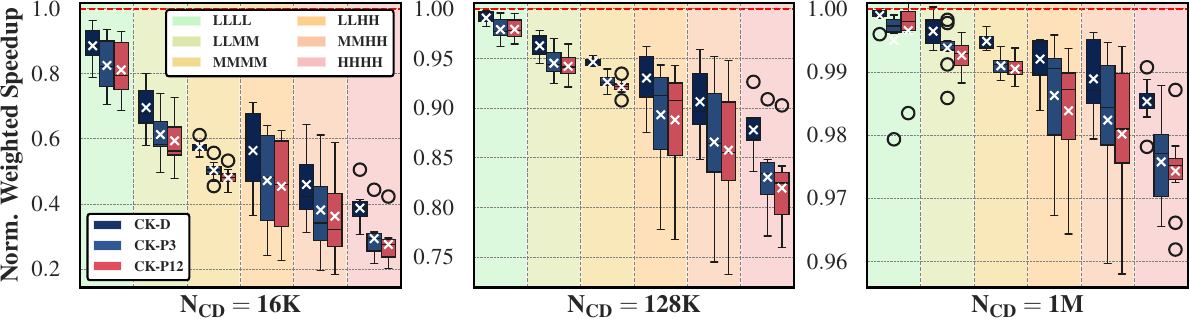}
    \caption{\revv{Performance impact of \papername{} on multi-core workloads for three different \CD{} thresholds.}}
    \label{fig:multi_core_perf}
\end{figure}

Figure~\ref{fig:multi_core_perf} presents \papername's performance impact on multi-core workloads.
The left, middle, and right subplots show results for $N_{CD}{=}16K$, $128K$, and $1M$, respectively.
Each boxplot corresponds to a particular \papername{} variant and workload mix category (denoted by background color), and shows the distribution of weighted speedup normalized to a baseline with no mitigation.

We make two key observations. 
First, \papername{} incurs higher performance overheads for multi-core workloads (compared to single-core).
For $N_{CD}{=}128K$, \CKD{} experiences a non-negligible average slowdown of \param{\SlMCAvgDMi\%}.
\CKP[3] and \CKP[12] experience a higher average slowdown of \param{\SlMCAvgPaMi\%} and \param{\SlMCAvgPbMi\%}, respectively.
Multi-core workloads generate more row-buffer conflicts, resulting in more preventive refreshes.
Second, the difference in performance between \CKD{} and \CKP{} is more pronounced compared to single-core workloads.
At a threshold of $128K$ ($16K$), \CKD{} \revv{has} \param{\SlMCDvsPbMi\%} (\param{\SlMCDvsPbLo\%}) higher normalized weighted speedup compared to \CKP[12].
In contrast, for single-core workloads, this difference is \param{\SlSCDvsPbMi\%} (\param{\SlSCDvsPbLo\%}).
We attribute this to the different applications in our multi-core workload mixes, which may frequently access neighboring subarrays, resulting in more instances of ``\textit{double-counting}'' (Figure~\ref{fig:double-count-example}), which \CKD{} accounts for but \CKP{} does not.

\subsubsection{Energy Consumption}
\begin{figure}[ht]
\centering   
    \includegraphics[width=\columnwidth]{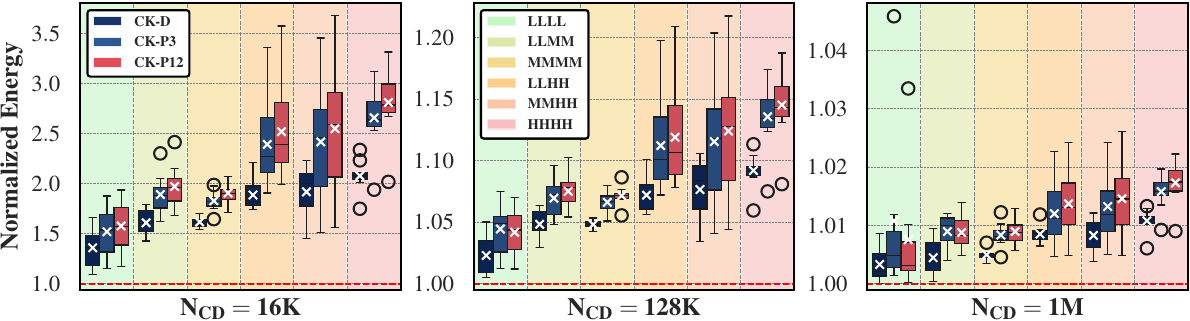}
    \caption{\revv{Effect of \papername{} on DRAM energy consumption for multi-core workloads.}}
    \label{fig:multi_core_energy}
\end{figure}

Figure~\ref{fig:multi_core_energy} presents \papername's impact on DRAM energy consumption for multi-core workloads.
The left, middle, and right subplots show results for $N_{CD}{=}16K$, $128K$, and $1M$, respectively.
Each boxplot corresponds to a particular \papername{} variant and workload mix category (denoted by background color) and shows the distribution of energy consumption normalized to a baseline with no \CD{} mitigation.
At $N_{CD}{=}1M$, all variants incur negligible energy consumption increases.
When $N_{CD}{=}128K$, \CKD{} increases average (maximum) energy consumption by \param{\EnMCAvgDMi\%} (\param{\EnMCMaxDMi\%}).
At this threshold, \CKP[3] (\CKP[12]) increases average energy consumption by \param{\EnMCAvgPaMi\%} (\param{\EnMCAvgPbMi\%}).
For $N_{CD}{=}16K$, all variants increase energy consumption significantly, with average values of \param{\EnMCAvgDLo$\times$}, \param{\EnMCAvgPaLo$\times$}, and \param{\EnMCAvgPbLo$\times$} over the baseline for \CKD{}, \CKP[3], and \CKP[12], respectively.

\subsection{Evaluation under Adversarial Workloads}
\label{sub:eval_adversarial}

\begin{figure}[ht]
\centering
        \includegraphics[width=\columnwidth]{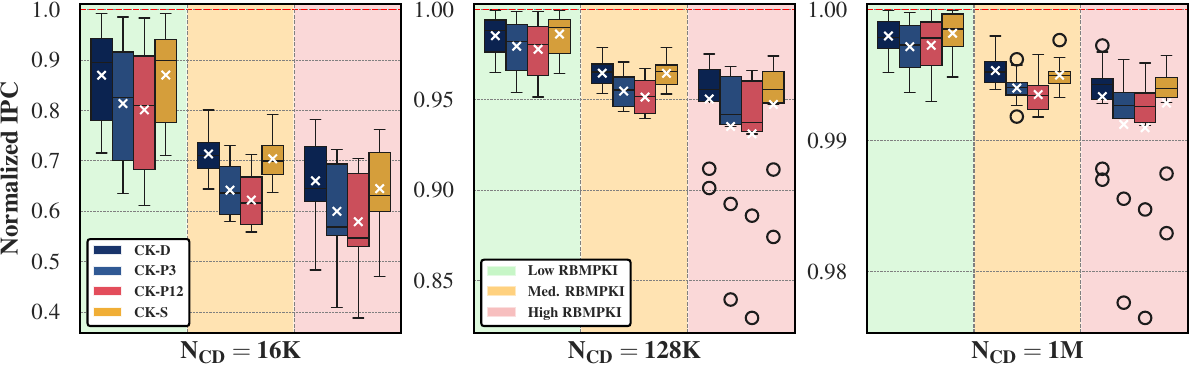}
        \caption{\revv{Performance impact of \papername{} on single-core workloads colocated with adversarial workloads.}}
        \label{fig:adv_eval}
\end{figure}

To evaluate \papername's performance under adversarial workloads, we colocate single-core workloads with a synthetic workload that rapidly hammers rows across three consecutive subarrays to perform a \CD{} attack.
Figure~\ref{fig:adv_eval} shows the performance impact of \CKD, \CKP, and \CKS{} (described in \secref{sec:methodology}) under this adversarial scenario.
The left, middle, and right subplots show results for $N_{CD}{=}16K$, $128K$, and $1M$, respectively.
Each individual boxplot corresponds to a particular \papername{} variant and RBMPKI category (denoted by background color), and shows the distribution of IPC, normalized to a baseline with no \CD{} mitigation.

\noindent We make three key observations. 
First, all mechanisms incur low average performance overheads for $N_{CD}{=}1M$ ($<$\param{\IpcAdvAllHiMaxOh\%}) or $128K$ ($<$\param{\IpcAdvAllMiMaxOh\%}).
Second, at low thresholds (i.e., $16K$), all mechanisms incur significant performance degradation ($\ge$\param{\IpcAdvAllLoMinOh\%}) due to the higher number of preventive refreshes issued.
Specifically, for \CKD, \CKS, and \CKP, when moving from a \CD{} threshold of $128K$ to $16K$, the number of preventive refreshes issued increases by \param{\MrAdvRatioDLoOverMi$\times$}, \param{\MrAdvRatioSLoOverMi$\times$}, and \param{\MrAdvRatioPLoOverMi$\times$}, respectively.
Third, the single-counter design of \CKS{} represents a middle ground between \CKD{} and \CKP{}, as it deterministically tracks activations, but does not account for potential \textit{double-counting}.
As a result, for $N_{CD}{=}16K$, \CKS{} issues \param{\MrAdvSvsDLo\%} (\param{\MrAdvSvsPbLo\%}) more (fewer) refreshes than \CKD{} (\CKP[12]), on average.
This induces an average IPC degradation of \param{\IpcAdvAvgDLo\%}, \param{\IpcAdvAvgSLo\%}, and \param{\IpcAdvAvgPbLo\%} for \CKD, \CKS, and \CKP[12], respectively.
These results \revv{are in accordance} with the observation made in \secref{motivation:double}, where \textit{double-counting} was defined.

\subsection{\RH{} \& \CD{}}
\label{sub:eval_rowhammer_cd}

\begin{figure}[ht]
\centering
    \includegraphics[width=\columnwidth]{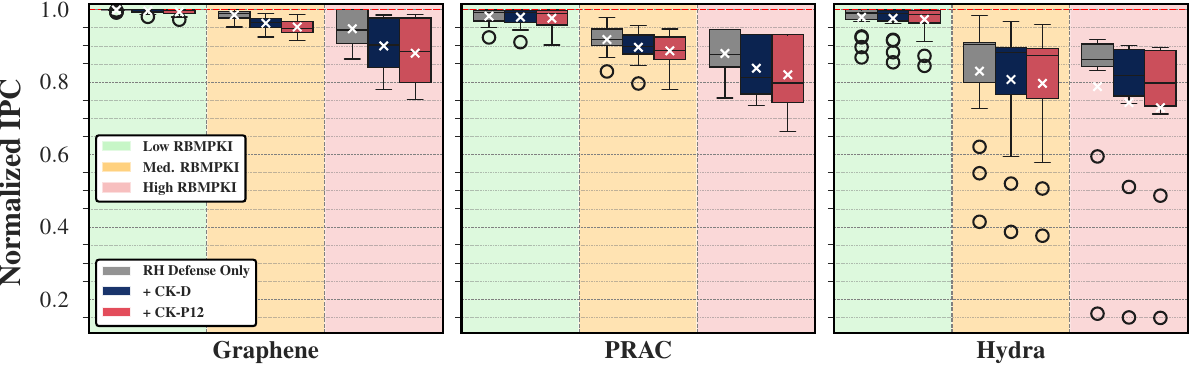}
    \caption{\revv{Performance impact of \papername{} with different \RH{} mitigations.} $\mathbf{N_{CD}{=}128K, N_{RH}{=}128}$.}
    \label{fig:cd_and_rh}
\end{figure}

Figure~\ref{fig:cd_and_rh} shows the performance impact of \papername{} when combined with three state-of-the-art \RH{} mitigation mechanisms: Graphene~\cite{park2020graphene}, PRAC~\pracCitations{}, and Hydra~\cite{qureshi2022hydra}.
Within each subplot, the background color denotes the RBMPKI category.
Each boxplot corresponds to a particular combination of \RH{} and \CD{} mitigation mechanisms and shows the distribution of IPC across single-core workloads, normalized to a baseline with no \RH{} or \CD{} mitigation.
All mitigation mechanisms operate transparently \revv{to each other}, taking each other's preventive refreshes into account (\secref{security:activations}).

\noindent We make the following key observations.
First, low-overhead \RH{} mitigations (e.g., Graphene~\cite{park2020graphene}) \revv{do not significantly increase} the performance overheads of \papername{}.
For example, \CKD{} + Graphene \revv{has} an average (minimum) normalized IPC of \param{\IpcRhGrapheneAvgD} (\param{\IpcRhGrapheneMinD}), compared to \param{\IpcRhCdOnlyAvgD} (\param{\IpcRhCdOnlyMinD}) \revv{for \CKD{} without Graphene} (\secref{eval:single}).
Second, \CKD{} (\CKP[12]) \revv{reduces normalized IPC by} \param{\SlRhPracAvgD\%} (\param{\SlRhPracAvgPb\%}) \emph{on top} of PRAC and \param{\SlRhHydraAvgD\%} (\param{\SlRhHydraAvgPb\%}) \emph{on top} of Hydra, on average.

\subsection{Sensitivity to DRAM Subarray Size}
\label{sub:sensitivity_study}

\begin{figure*}[bt]
\centering
    \includegraphics[width=\textwidth]{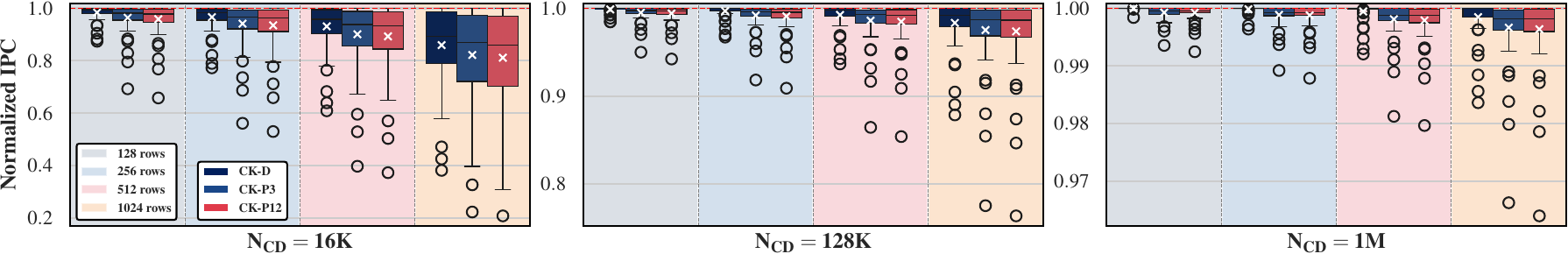}
    \caption{\revv{Performance impact of \papername{} across different subarray sizes for $\mathbf{N_{CD}{=}16K}$, $\mathbf{128K}$, and $\mathbf{1M}$.}}
    \label{fig:subarray_size_analysis}
\end{figure*}

Figure~\ref{fig:subarray_size_analysis} shows the impact of subarray size ($S$) on single-core performance with \papername{}.
The three subplots show results for $N_{CD}{=}16K$ (left), $128K$ (middle), and $1M$ (right).
We evaluate subarray sizes of $128$, $256$, $512$, and $1K$ rows (denoted by different background colors).
Each boxplot corresponds to a particular \papername{} variant and subarray size, and shows the distribution of normalized IPC across all workloads, \revv{with respect to} a baseline with no mitigation.

We observe that across all evaluated thresholds, the performance \revv{degradation} of all mechanisms decreases significantly with smaller subarray sizes.
This is expected from \secref{security:d} and \secref{security:p}, as the \revv{subarray size} $S$ is inversely (positively) correlated with the \textit{preventive refresh threshold}, $N_{PR}$ of \CKD{} (\textit{preventive refresh probability}, $P_{PR}$ of \CKP{}).
For $N_{CD}{=}128K$, when reducing the subarray size from $1K$ rows (as reported in~\cite{columndisturb}) to $256$ rows, the average IPC degradation for \CKD{} and \CKP[12] drops from \param{\SlSubAvgDOneKMi\%} to \param{\SlSubAvgDTwoFiveSixMi\%}, and from \param{\SlSubAvgPbOneKMi\%} to \param{\SlSubAvgPbTwoFiveSixMi\%}, respectively.
With smaller subarrays, \papername{} incurs \revv{relatively low} overheads even at very low thresholds.
For example, when $S{=}128$ and $N_{CD}{=}16K$, \CKD{} (\CKP[12]) incurs an average slowdown of \param{\SlSubAvgDSmallLo\%} (\param{\SlSubAvgPbSmallLo\%}) and a maximum of \param{\SlSubMaxDSmallLo\%} (\param{\SlSubMaxPbSmallLo\%}).

\subsection{Sensitivity to Physical Page Allocation}

\begin{figure}[ht]
\centering
    \includegraphics[width=\columnwidth]{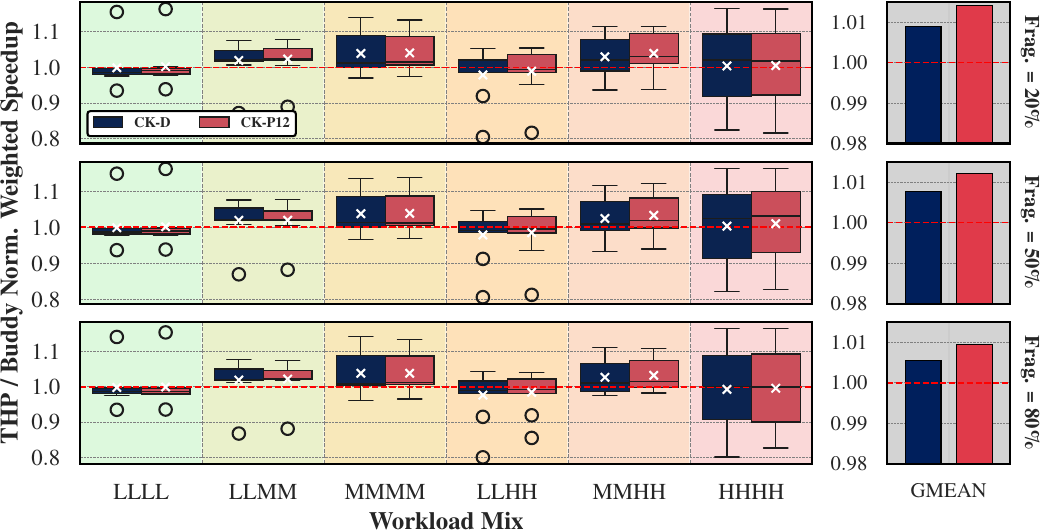}
    \caption{Weighted speedup of THP normalized to the Buddy allocator for different fragmentation levels at $\mathbf{N_{CD}{=}128K}$.}
    \label{fig:allocator}
\end{figure}

To understand the impact of physical page allocation on \papername's performance, we evaluate \CKD{} and \CKP[12] on multi-core workload mixes with: (i) the standard 4KB Buddy allocator~\cite{buddy}, and (ii) a Transparent Huge Pages (THP)-like allocator~\cite{corbet2011,arcangeli2010} that attempts to allocate 2MB pages and falls back to 4KB pages.
To decouple the performance impact of page allocation from address translation, we use zero-overhead address translation.
Figure~\ref{fig:allocator} shows \papername's performance impact with both allocators at $N_{CD}{=}128K$.
The top, middle, and bottom rows show results for memory fragmentation levels of $20\%$, $50\%$, and $80\%$ at \texttt{2MB} granularity \revv{(i.e., the ratio of free memory that \emph{cannot} be allocated as \texttt{2MB}-aligned contiguous blocks)}, respectively.
Each boxplot (left) corresponds to a particular workload mix category (denoted by background color), and shows the distribution of weighted speedup when using THP normalized to Buddy.
Each barplot (right) shows the geomean across all workload mixes.

\noindent We observe that there is no significant performance difference between the two allocators at any fragmentation level.
Specifically, for \CKD{} (\CKP[12]), the geomean normalized weighted speedup is \param{\IpcFragGmDTwenty} (\param{\IpcFragGmPbTwenty}), \param{\IpcFragGmDFifty} (\param{\IpcFragGmPbFifty}), and \param{\IpcFragGmDEighty} (\param{\IpcFragGmPbEighty}) for fragmentation levels of $20\%$, $50\%$, and $80\%$, respectively.

\subsection{Evaluation with Subarray-Level Parallelism}
\label{eval:salp}
\begin{figure}[ht]
\centering
    \includegraphics[width=\columnwidth]{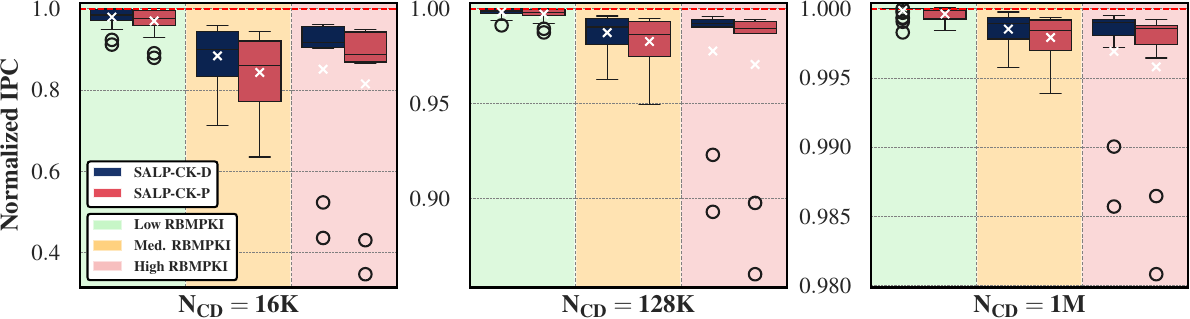}
    \caption{Performance impact of \papername{} in a system that employs MASA~\cite{kim2012case}.}
    \label{fig:salp_eval}
\end{figure}

Subarray-level parallelism (SALP)~\cite{kim2012case} is a \revv{DRAM microarchitecture modification that allows multiple subarrays (i.e., multiple row buffers) within the same bank to be activated simultaneously.}
\revv{SALP~\salpCitations{} improves performance by reducing the impact of bank conflicts.
For \papername, SALP allows preventive refreshes in one subarray to overlap with DRAM requests in other subarrays.}
Figure~\ref{fig:salp_eval} shows \papername's performance impact on a system that uses the MASA~\cite{kim2012case} SALP configuration.
The left, middle, and right subplots show results for $N_{CD}{=}16K$, $128K$, and $1M$, respectively.
Each boxplot corresponds to a particular MASA-enabled \papername{} mechanism and RBMPKI category (denoted by background color), and shows the distribution of IPC across all single-core workloads, normalized to a MASA-enabled system with no \CD{} mitigation.

We make three key observations.
First, at $N_{CD}{=}1M$, \CKDSALP{} and \CKPSALP{} incur negligible average (maximum) performance overheads of \param{\SlSalpAvgDHi\%} (\param{\SlSalpMaxDHi\%}) and \param{\SlSalpAvgPbHi\%} (\param{\SlSalpMaxPbHi\%}), respectively.
Second, for $N_{CD}{=}128K$, the overheads remain low, with average (maximum) values of \param{\SlSalpAvgDMi\%} (\param{\SlSalpMaxDMi\%}) for \CKDSALP{} and \param{\SlSalpAvgPbMi\%} (\param{\SlSalpMaxPbMi\%}) for \CKPSALP{}.
Third, even at the very low threshold of $16K$, \CKDSALP{} (\CKPSALP{}) mitigates \CD{} with an average overhead of \param{\SlSalpAvgDLo\%} (\param{\SlSalpAvgPbLo\%}) and a maximum slowdown of \param{\SlSalpMaxDLo\%} (\param{\SlSalpMaxPbLo\%}).
This is substantially lower than the maximum slowdowns of over \param{\SlNoSalpMaxDLo\%} and \param{\SlNoSalpMaxPbLo\%} observed for the non-SALP \CKD{} and \CKP[12] configurations (\secref{eval:single}).
We conclude that SALP alleviates \papername's performance overheads.

%% file: sections/07_implementation.tex
\section{Latency, Area \& Power Overheads}
\label{sec:implementation}

\CGUARDD{} (\CKD) and \CGUARDP{} (\CKP) are implemented in the memory controller and require no modifications to DRAM.
Each \CTO{} and \CTE{} entry contains a maximum value of $(N_{CD}{-}2S)/S{\approx} N_{CD}/S$, \revv{where $S$ is the subarray size}.
Each \RPT{} entry points to one of $S$ rows in a subarray.
As a result, each \CTO{} and \CTE{} entry requires $\log_2(N_{CD}{/}S)$ bits of storage, and each \RPT{} entry requires $\log_2(S)$ bits of storage.
Given: (i)~a \CD{} threshold ($N_{CD}$) of $1M$, (ii)~a subarray size $S{=}1K$, and (iii)~the dual-rank configuration of \secref{sec:methodology}, the total storage overheads of \CGUARDD{} (\CGUARDP{}) \revv{are} $7.5$KB ($2.5$KB).
We implement \CKD{} and \CKP{} in Verilog HDL and use open-source tools~\cite{openroad,nangate45} to synthesize them and evaluate their latency, area, and power overheads.
\CKD{} (\CKP) increases the processor's power consumption and area by \param{50\,mW} (\param{15\,mW}) and \param{0.1\,mm$^2$} (\param{0.03\,mm$^2$}), respectively.
The row-to-row activation delay ($t_{RRD}$) determines the minimum latency between successive row activations across banks, with $t_{RRD}{\ge}2.5$\,ns in DDR4.
\CKD{} (\CKP{}) can read and update the necessary counters while meeting this timing constraint, with a slack of \param{0.7\,ns} (\param{0.5\,ns}).
As these operations happen off the critical path of the memory controller, they add no additional memory request latency.

%% file: sections/08_indram.tex
\section{Alternative In-DRAM Implementation}
\label{disc:indram}

\papername's memory-controller-based design eases its adoption by avoiding changes to the density-optimized DRAM chips.
However, this design choice introduces two drawbacks.
First, it requires exposing (or reverse-engineering, \secref{disc:reveng}) the subarray mapping to the memory controller.
Second, as the memory controller \emph{cannot} determine which subarrays \texttt{RFM}~\cite{canpolat2025chronus,jedecddr5c} affects, it conservatively increments multiple counters when issuing such commands, increasing \papername's overheads.

To alleviate these issues, we explore a potential in-DRAM design of \CKD, which reuses PRAC's~\pracCitations{} standardized \texttt{ALERT} signal and \texttt{RFM} command.
We \revv{place} the \RPT{}, \CTO{}, and \CTE{} entries next to each subarray and introduce two registers in each bank \revv{to track the subarray with the highest activation count}: (i) a \textbf{Subarray Pointer} (SP), which points to the subarray in the bank with the highest activation count, and (ii) a \textbf{Subarray Hammer Counter} (SHC), which contains the maximum value of \CTO{} and \CTE{} for the subarray pointed to by SP.
Each time \CKD{} updates its counter entries, it checks whether the new values exceed SHC and updates it accordingly.
When SHC exceeds $N_{PR}$, \CKD{} asserts the \texttt{ALERT} signal, which informs the memory controller to issue an \texttt{RFM} command to the entire DRAM rank.
When a bank receives \texttt{RFM}, it first identifies the subarray with the highest activation count (via SP).
Second, it refreshes the next \param{7} rows pointed to by \RPT{}.\footnote{\texttt{RFM} commands stall every bank in an entire rank for \param{350\,ns}. This provides enough time to refresh \param{7} rows in each bank with one \texttt{RFM} command~\cite{qureshi2026salt}.}
Third, it increments \RPT{} (by \param{7}) and resets \CTO{} and \CTE{} to \param{7}.
Fourth, it recalculates SP and SHC with the updated \CTO{} and \CTE{} entries in the neighboring subarrays.

\begin{figure}[ht]
\centering
    \includegraphics[width=\columnwidth]{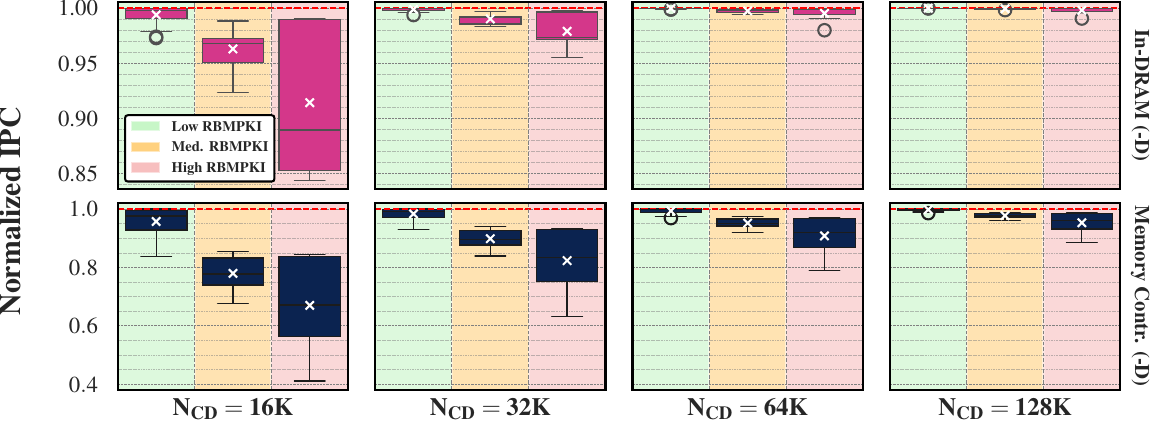}
    \caption{\revv{Performance impact of \CKD{} (bottom) and its in-DRAM implementation (top) with PRAC enabled, $\mathbf{N_{RH}{=}128}$.}}
    \label{fig:in-dram-evaluation}
\end{figure}

Figure~\ref{fig:in-dram-evaluation} presents the performance impact of the in-DRAM and the memory-controller-based implementations of \CKD{} on single-core workloads, when integrated with PRAC ($N_{RH}{=}128$).
The top (bottom) row shows results for the in-DRAM (memory-controller-based) implementation.
From left to right, each column shows results for $N_{CD}{=}16K$, $32K$, $64K$, and $128K$, respectively.
Each boxplot corresponds to a particular RBMPKI category (denoted by background color) and shows the distribution of IPC, normalized to a PRAC-enabled system with no \CD{} mitigation.

\noindent First, we observe that for $N_{CD}{=}128K$, the in-DRAM version \revv{of \CKD{}} incurs negligible IPC degradation, with a minimum normalized IPC of \param{\DiscIpcInDramMinMi}, compared to \param{\DiscIpcMemCtrMinLo} for the memory-controller-based implementation.
Second, for $N_{CD}{=}16K$, where the memory-controller-based implementation has an average (minimum) normalized IPC of \param{\DiscIpcMemCtrAvgLo} (\param{\DiscIpcMemCtrMinLo}), the in-DRAM version maintains a high average (minimum) IPC of \param{\DiscIpcInDramAvgLo} (\param{\DiscIpcInDramMinLo}).
\revv{This is because the in-DRAM implementation can identify which subarrays are affected by RowHammer-preventing \texttt{RFM} commands, whereas the memory-controller-based implementation conservatively increments all subarray counters in the entire rank, leading to unnecessary refreshes and higher performance overheads.}
We conclude that an in-DRAM implementation of \CKD{} can provide low-overhead \CD{} mitigation at very low thresholds.

%% file: sections/09_discussion.tex
\section{Discussion}

\subsection{Subarray Mapping Reverse-Engineering}
\label{disc:reveng}

\papername{} requires knowledge of DRAM's subarray mapping.
Even if this mapping is not made available by the DRAM manufacturer, it can still be reverse-engineered by first clustering rows into subarrays and then discovering adjacent subarrays.
The proposed reverse-engineering methods can be triggered from software~\cite{kim2014flipping,luo2023rowpress,loughlin2023siloz}, or via a programmable FPGA-based memory controller~\cite{safari-drambender,softmcgithub}.

\noindent\textbf{Clustering Rows Into Subarrays.}
Multiple prior works \cite{kim2020revisiting,yaglikci2024svard,hassan2021utrr,orosa2021deeper,columndisturb} propose methods to reverse-engineer DRAM subarray mappings and boundaries.
Sv{\"a}rd~\cite{yaglikci2024svard} examines two approaches.
The first uses RowHammer~\RHFundamentalCitations{} to cause bitflips in physically adjacent rows and employs k-means clustering to group them into subarrays.
The second (also used in~\cite{yuksel2024functionally,columndisturb,nam2024dramscope}) uses RowClone's~\cite{seshadri2013rowclone} copy operation, which copies one row to another row in the same subarray.
This operation fails for rows in different subarrays, so~\cite{yaglikci2024svard} tests all row pairs and discovers which pairs reside in the same subarray.

\noindent\textbf{Discovering Adjacent Subarrays.}
Following the previous step, we can reverse-engineer the adjacency of the discovered subarrays using \CD~\cite{columndisturb} or FC-DRAM~\cite{yuksel2024functionally}.
\underline{\CD~\cite{columndisturb}} hammers a row in a subarray to induce bitflips in that subarray or its neighbors.
We (i) initialize DRAM to a known data pattern, (ii) repeatedly hammer a subarray enough times ($N_{CD}$) to cause bitflips, and (iii) read out the contents of memory.
Any rows containing cells with bitflips must reside either in the hammered subarray (already known) or its immediate neighbors.
By repeating this for all subarrays, we can uncover the adjacency between them.
\underline{FC-DRAM~\cite{yuksel2024functionally}} leverages the open-bitline DRAM architecture to write the negated value of a DRAM row in one subarray to a row in a neighboring subarray.
We (i) initialize DRAM to a known data pattern, (ii) randomly select a row in a subarray, and (iii) perform the NOT operation with a row in a different subarray as the destination.
By evaluating \revv{whether or not the} NOT operation \revv{succeeds}, we detect whether two subarrays are adjacent. We repeat this until we uncover the adjacency of all subarrays.

While the above methods are reliable,\footnote{Up to 99.9\% of DRAM rows are RowHammer-vulnerable~\cite{hassan2021utrr}. \emph{All} rows across 3 subarrays experience \CD{} bitflips~\cite{columndisturb}. Uncovering adjacency requires discovering \revv{only} a single row containing a bitflip.} they rely on read disturbance~\cite{kim2014flipping,luo2023rowpress,columndisturb} or timing parameter violations~\cite{seshadri2013rowclone,yuksel2024functionally}.
To keep \CKD{} secure, when a row cannot be mapped to a subarray, we can increment the counters for all subarrays in the row's bank (as done for \texttt{RFM} in \secref{security:activations}). The \CKP{} equivalent is ``\textit{flipping a coin}'' for all subarrays in the same bank.

\subsection{DRAM Microarchitecture Modifications}

\secref{eval:salp} and \secref{sub:sensitivity_study} discussed changes to DRAM microarchitecture such as SALP~\salpCitations{} and smaller subarrays to reduce \papername's performance impact.
Such changes introduce performance, power, and area tradeoffs.
For example, SALP~\cite{kim2012case} improves single-core IPC by \param{17\%} but increases DRAM area by \param{0.15\%} and power consumption by \param{0.56\,mW} (per concurrently activated subarray)~\cite{kim2012case}.
Designs that employ smaller or asymmetric subarrays (e.g.,~\cite{lee2013tiered,lu2015improving,luo2020clrdram,kim2012case}) offer lower DRAM latency and energy consumption by reducing bitline capacitance.
However, they increase area (e.g., \revv{by} \param{6.6\%}~\cite{lu2015improving}) \revv{as they introduce} new DRAM peripherals.

An alternative way to reduce the impact of \CD{} and \papername's overheads is to introduce isolation transistors that disconnect the bitlines of neighboring subarrays from the sense amplifiers.
Apart from preventing hammers from affecting neighboring subarrays, isolation transistors enable (i) inter-subarray row copy operations~\cite{chang2016lisa}, (ii) lower DRAM latencies~\cite{seongil2014row,lee2013tiered}, and (iii) RowPress~\cite{luo2023rowpress} mitigations~\cite{qureshi2025moat,marazzi2023rega}.
Both~\cite{chang2016lisa} and~\cite{seongil2014row} report that isolation transistors introduce a \param{0.8\%} DRAM area overhead.

%% file: sections/10_related.tex
\section{Related Work}
\label{sec:related}

\revv{To our knowledge, this is the first work to design and comprehensively evaluate mechanisms that mitigate \CD{} bitflips at current and future thresholds.
We propose \CGUARDD{} (\CKD), a new deterministic mitigation mechanism with low performance and energy overhead, and \CGUARDP{} (\CKP), a new probabilistic mechanism at lower area overhead.}

\revv{While the \CD{} paper~\cite{columndisturb} \emph{sketches} a mechanism similar to \CKD{}, we propose, design, implement and comprehensively analyze the security of \CKD{} and \CKP{}, and evaluate their performance, energy, power, and area overheads.
A concurrent work~\cite{qureshi2026salt} proposes tracking row activations in-DRAM at subarray granularity to mitigate \RH{} and also describes possible modifications to mitigate \CD{}.
Silver Bullet~\cite{devaux2021method,yaglikci2021security} is a \RH{} mitigation mechanism employed in UPMEM systems~\cite{devaux2019true,gomez2022benchmarking,gomez2023evaluating,gomez2021benchmarkingcut} that tracks row activations at ``\revv{subbank}'' granularity.}
\revv{Similarly to the \RPT{} in \CKD{} and \CKP{}, Silver Bullet~\cite{devaux2021method,yaglikci2021security} employs a single pointer per subbank to iteratively refresh all rows in a subbank.}
Silver Bullet~\cite{devaux2021method,yaglikci2021security} is not designed to protect against ColumnDisturb but it can be adapted for that purpose.
This section discusses other relevant \RH{} mitigations mechanisms.

\noindent\textbf{Per-Row Activation Tracking.}
Various prior works leverage per-row activation counters to track the number of DRAM row activations within a refresh interval~\cite{kim2014flipping,bennett2021panopticon,kim2014architectural,kim2023ddr5,yaglikci2021security}.
An update to the JEDEC DDR5 specification~\cite{jedecddr5c,saroiu2024ddr5} introduces a similar on-DRAM-die read disturbance mitigation framework called Per Row Activation Tracking (PRAC)~\pracCitations.
PRAC aims to ensure robust operation and mitigate \RH{} bitflips with low performance overhead by deterministically tracking the activations of each row and preventively refreshing victim rows when necessary.
However, as discussed in \secref{motivation}, PRAC is \CD-oblivious and \emph{cannot} mitigate \CD{} bitflips (at low overhead).
\revv{In \secref{disc:indram}, we show that by combining PRAC with a potential in-DRAM implementation of \CKD{}, it is possible to mitigate \CD{} bitflips with low overheads even at very low thresholds.}

\noindent\textbf{Other On-Die \RH{} Mitigation Mechanisms.}
DRAM manufacturers implement \RH{} mitigation mechanisms, such as Target Row Refresh (TRR), in commercial DRAM chips~\cite{jedec2020ddr5,jedec2017ddr4}.
However, custom attacks can bypass these mechanisms~\cite{frigo2020trrespass,hassan2021utrr,jattke2022blacksmith,deridder2021smash,van2016drammer,saroiu2022price} and cause \RH{} bitflips.
To adopt \papername{} in a system with TRR, \papername{} needs to track target row refreshes.
Similar to \RH{} mitigation proposals implemented in the memory controller, \papername{} requires information on how TRR works and which rows it refreshes to accurately track activation counts.
This information can be provided to \papername{} to securely mitigate \CD{}.

\newpage

\noindent\textbf{Memory Controller-Based \RH{} Mitigations.}
Various prior works propose memory controller-based \RH{} mitigations.
These mechanisms perform preventive refreshes either probabilistically~\cite{kim2014flipping,yaglikci2022hira,you2019mrloc,seyedzadeh2017cbt}, or deterministically by tracking DRAM row activations~\deterministicRowHammerDefenseCitations.
A subset of these works leverages frequent item counting algorithms such as Misra-Gries~\cite{olgun2024abacus,park2020graphene} or Count-Min-Sketch~\cite{bostanci2024comet} to reduce the number of activation counters and thus the area overhead of activation count tracking.
\papername{} is a memory controller-based \CD{} mitigation mechanism that can be employed together with these mechanisms to prevent both \CD{} and \RH{} bitflips.
In contrast to the trackers of these deterministic \RH{} mitigations, \papername{} only needs to track activations at the subarray level, hence requiring less area (see \secref{sec:implementation}) for activation counters.

\noindent\textbf{\revv{Exploiting} Subarray-Level Parallelism (SALP) in Read Disturbance Mitigations.}
\revv{Prior works~\cite{marazzi2023rega,yaglikci2022hira} leverage subarray-level parallelism~\cite{kim2012case} to reduce the overheads of \RH{} mitigations.
SALP~\salpCitations{} reduces the performance impact of bank conflicts induced by \RH{} preventive refreshes.
REGA~\cite{marazzi2023rega} introduces a second sense amplifier in each subarray that can be used to overlap preventive refresh operations with regular memory accesses to the same subarray.
HiRA~\cite{yaglikci2022hira} parallelizes preventive refresh operations in one subarray with regular memory accesses to other subarrays within the same bank by carefully violating timing parameters in real DRAM chips.
These approaches are orthogonal to ours and can be combined with \papername{} to reduce its performance overheads (as also shown by our evaluation with SALP~\salpCitations{} in \secref{eval:salp}).}

%% file: sections/11_conclusion.tex
\section{Conclusion}
\label{sec:conclusion}

We introduced the first mitigation mechanisms that protect against \CD{} bitflips \revv{in DRAM chips} either deterministically (\CGUARDD), with reduced performance and energy overheads, or probabilistically (\CGUARDP), with lower hardware complexity.
Both mechanisms protect against \CD{} bitflips with low performance and energy overheads at current ($1M$) and near-future ($128K$) thresholds.
Handling lower thresholds (e.g., $16K$) with low overheads is possible with modifications to DRAM microarchitecture, or with an in-DRAM implementation of our deterministic mechanism.
\revv{We hope future work builds on ours and develops other solutions to \CD{}.}
\revv{\papername{} is freely available at \href{https://github.com/CMU-SAFARI/ColumnKeeper}{github.com/CMU-SAFARI/ColumnKeeper}}.

%% file: sections/12_acknowledgements.tex
\section*{Acknowledgments}

We thank the anonymous reviewers and artifact evaluators of ISCA 2026 for feedback.
We thank the SAFARI Research Group members for their constructive feedback and for providing a stimulating intellectual environment. 
We acknowledge the generous gift funding provided by our industrial partners (especially Google, Huawei, Intel, Microsoft), which has been instrumental in enabling the research we have been conducting on read disturbance in DRAM in particular and memory systems in general~\cite{mutlu2017rowhammer,mutlu2019processing,mutlu2019rowhammer,mutlu2022modern,mutlu2023fundamentally,mutlu2023retrospective,mutlu2023retrospective-isca,mutlu2014research,mutlu2023retrospective-raidr,mutlu2025memory,mutlu2013memory}.
This work was in part supported by a Google Security and Privacy Research Award and the Microsoft Swiss Joint Research Center.

%% file: sections/13_artifact_appendix.tex
\appendix
\section{Artifact Appendix}

\subsection{Abstract}

This artifact provides the infrastructure to reproduce \papername{}'s main experimental results. It includes \papername's integration into Ramulator 2.0~\ramulatorCitations, a \texttt{C++}-based Monte Carlo simulation, 62 workload traces from 5 benchmark suites (SPEC CPU2006, SPEC CPU2017, TPC, MediaBench, and YCSB), and automated scripts for running all experiments and generating Figures 2 and 5--15 of the paper.
The artifact evaluates \papername{}'s impact on system performance and DRAM energy consumption across a range of \CD{} thresholds, subarray sizes, and system configurations.

\subsection{Artifact check-list (meta-information)}

\begin{itemize}
    \item \textbf{Program:} Ramulator 2.0 cycle-accurate DRAM simulator with \papername{} mitigation implementations, and a Monte Carlo simulation framework.
    \item \textbf{Compilation:} C++ with GCC (tested on 11.3.0), CMake-based build.
    \item \textbf{Data set:} 62 workload traces from SPEC CPU2006, SPEC CPU2017, TPC, MediaBench, and YCSB; 1 synthetic adversarial trace auto-generated during installation.
    \item \textbf{Run-time environment:} Linux (tested on Ubuntu 22.04.01 LTS), Python 3.10, Conda 24.11.1.
    \item \textbf{Hardware:} x86\_64-based compute cluster.
    \item \textbf{Execution:} Trace-driven simulation, Slurm job scheduler supported.
    \item \textbf{Metrics:} Normalized IPC (Instructions Per Cycle), weighted speedup, normalized DRAM energy consumption.
    \item \textbf{Output:} PDF visualization plots.
    \item \textbf{How much time is needed to complete experiments (approximately)?:} $\sim$12 hours using 300 CPUs.
    \item \textbf{Publicly available?:} Yes.
    \item \textbf{Workflow automation framework used?:} Python, Bash, and Slurm job scheduler.
    \item \textbf{Archived (DOI):} \url{https://doi.org/10.5281/zenodo.19446517}
\end{itemize}

\subsection{Description}

\subsubsection{How to Access}
The artifact (workload traces and code) is available on Zenodo via the following link: \url{https://doi.org/10.5281/zenodo.19446517}

\subsubsection{Hardware Dependencies}
x86\_64 compute cluster.

\subsubsection{Software Dependencies}
\begin{itemize}
    \item Linux (tested on Ubuntu 22.04.01 LTS)
    \item Slurm Workload Manager (tested on slurm-wlm 21.08.5)
    \item Conda (tested on 24.11.1)
    \item GCC (tested on 11.3.0)
    \item CMake (tested on 3.22.1)
    \item Python 3.10 (conda-managed installation and packages)
    \item pdfTeX (tested on 3.141592653-2.6-1.40.29, TeX Live 2026)
\end{itemize}

\subsubsection{Data sets}
The artifact uses 62 traces from 5 benchmark suites: (i) SPEC CPU2006, (ii) SPEC CPU2017, (iii) TPC, (iv) MediaBench, and (v) YCSB, along with 1 synthetic adversarial trace that is auto-generated during installation.

\subsection{Installation}

Assuming that all hardware and software dependencies are met, run \texttt{bash -i setup.sh} in the \texttt{ColumnKeeper/} directory.
The setup script will prompt the user to specify: (i) the path to the directory hosting the artifact code (downloaded from Zenodo), (ii) the path to \texttt{cputraces.tar.gz} (also downloaded from Zenodo), and (iii) the maximum number of concurrent experiments to run via Slurm (\texttt{MAX\_CONCURRENT\_JOBS}).
Upon completion, the script compiles the necessary executables (\texttt{ramulator2} for the performance evaluation and \texttt{mc\_worker} for the Monte Carlo simulations), decompresses the workload traces, and generates the synthetic traces.

\subsection{Experiment workflow}

After installation, the script continues its execution.
It first performs profiling runs on the workload traces to measure baseline results, then performs the Monte Carlo simulation for Figure 5, and finally runs the performance evaluation on Ramulator 2.0.
The script automatically: (i) creates the experimental configurations, (ii) schedules the configurations via Slurm, (iii) checkpoints its progress, and (iv) reschedules experiments that fail or time out.
If at any point the installation or execution is interrupted, restart from the last checkpoint by again running: \texttt{bash -i setup.sh}.

\subsection{Expected Results}

Upon completion, the artifact produces reproduced versions of Figures 2 and 5--15 of the paper.
All generated figures are placed in the \texttt{figures/} directory within the \texttt{ColumnKeeper/} directory.
The setup script also autogenerates the \texttt{eval\_params.tex} file within the \texttt{paper/} directory, which contains \LaTeX{} macros for the numerical values of the performance and energy results for all evaluated configurations.
These macros can be used to automatically update the paper's key numerical results by recompiling the paper with \texttt{make} in the \texttt{paper/} directory.

\subsection{Manual Execution}

The \texttt{E*.py} files in the \texttt{experiments/} directory create the experimental configurations.
The \texttt{P*.py} files in the \texttt{plotting/} directory plot the figures.
Each \texttt{E*.py} file creates a directory with the experimental configurations in \texttt{experiments/results/}.
The user can either manually run these configurations with \texttt{ramulator2}, or use \texttt{sbatch} to schedule batches of jobs (using the batch files ending in \texttt{*\_b*.sh}).
Running \texttt{python -m plotting.G0\_eval\_params} after the experiments complete regenerates the numerical values cited in Sections 6 and 8 of the paper.
For more information, refer to the \texttt{README.md} file.